\documentclass[10pt,conference]{IEEEtran}
\usepackage{cite}
\usepackage{amsmath,amssymb,amsfonts}
\usepackage{algorithmic}
\usepackage{graphicx}
\usepackage{textcomp}
\usepackage{xcolor}
\usepackage[hyphens]{url}
\usepackage{fancyhdr}
\usepackage{hyperref}
\usepackage{verbatim}
\usepackage{adjustbox}
\usepackage{tcolorbox}
\usepackage{enumitem}
\usepackage{soul, xcolor}

\setulcolor{red}
\setstcolor{red}

\newcommand{\hpcayear}{2027}
\newcommand\equalcontrib{\textsuperscript{\textdagger}}

  

\newcommand{\hpcasubmissionnumber}{XXX}
\title{Starscream: Reimagining LLM Sharding on AMD Instinct GPUs with GPU Partitioning\vspace{-4mm}}

\def\hpcacameraready{} 
\newcommand\hpcaauthors{Sumanth Gudaparthi\equalcontrib, Smeet Chheda\equalcontrib, and Ganesh Dasika}
\newcommand\hpcaaffiliation{AMD Research and Advanced Development}
\newcommand\hpcaemail{\texttt{\{sumanth.gudaparthi, smeet.chheda, ganesh.dasika\}@amd.com}}

\author{
  \ifdefined\hpcacameraready
  \IEEEauthorblockN{\hpcaauthors{}}
      \IEEEauthorblockA{
        \hpcaaffiliation{} \\
        \hpcaemail{}
      }
  \else
    \IEEEauthorblockN{\normalsize{HPCA \hpcayear{} Submission
      \textbf{\#\hpcasubmissionnumber{}}} \\
      \IEEEauthorblockA{
        Confidential Draft \\
        Do NOT Distribute!!
      }
    }
  \fi 
}

\fancypagestyle{camerareadyfirstpage}{%
  \fancyhead{}
  
  \fancyhead[C]{
    \ifdefined\aeopen
    \else
      \ifdefined\aereviewed
      \else
      \ifdefined\aereproduced
      \else
    \fi 
    \fi 
    \fi 
    \ifdefined\aeopen 
      \includegraphics[width=12mm,height=12mm]{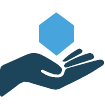}
    \fi 
    \ifdefined\aereviewed
      \includegraphics[width=12mm,height=12mm]{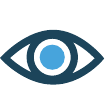}
    \fi 
    \ifdefined\aereproduced
      \includegraphics[width=12mm,height=12mm]{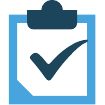}
    \fi
  }
  \fancyfoot[C]{}
}
\begin{document}
\maketitle

  \ifdefined\hpcacameraready
    \renewcommand\thefootnote{\textdagger}%
    \footnotetext{These authors contributed equally to this work.}%
    \renewcommand\thefootnote{\arabic{footnote}}%
  \fi
\ifdefined\hpcacameraready 
  \thispagestyle{camerareadyfirstpage}
  \pagestyle{empty}
\else
  \thispagestyle{plain}
  \pagestyle{plain}
\fi

\newcommand{\hpcaheight}{0mm}
\ifdefined\eaopen
\renewcommand{\hpcaheight}{12mm}
\fi


\begin{abstract}
As GPU accelerators grow larger to meet rising compute demands, they exhibit increasingly pronounced non-uniform memory access (NUMA) characteristics that challenge existing LLM
  frameworks. Chiplet-based architectures provide a practical path to larger GPUs while mitigating
  manufacturing yield constraints. Modern GPUs such as AMD Instinct MI300X employs such a multi-die design, and further offer \textit{GPU partitioning}\footnote{not to be confused with sharding: LLM model split across multiple devices} capabilities---partitioning a single physical GPU into multiple logical devices with NUMA-aware memory domains---promising faster kernel launches, and reduced memory access latencies with improved locality. However, current parallelization strategies, designed for homogeneous GPU topologies, when applied under partitioning, extend its unidimensional sharding causing significant performance degradation. Three bottlenecks emerge: (1) memory redundancy from replicated parameters or KV caches; (2) arithmetic intensity collapse as matrices become skinny and memory-bound; (3) communication overhead from flat collectives ignoring the 10$\times$ intra-GPU versus inter-GPU bandwidth difference. We present Starscream, a rethinking of LLM sharding for the chiplet era. Rather than stretching inter-GPU sharding thinner, Starscream adds intra-GPU sharding along an orthogonal dimension, grounded in three foundational principles: kernel shape preservation, memory redundancy
  elimination, and communication overhead minimization. These principles form the basis for extending any sharding strategy into the partitioning regime. We demonstrate their effectiveness using tensor parallelism as a case study, yielding sequence-length sharded attention eliminating KV replication, two-dimensional GEMM sharding maintaining arithmetic intensity, hierarchical collectives exploiting heterogeneous links, and computation-communication fusion via symmetric memory. On AMD MI300X and MI355X, exploiting partitioning capabilities yields up to 4× speedup for attention and 1.43× for fused GEMMs compared to chiplet-oblivious baselines, demonstrating the effectiveness of chiplet-aware designs. 
Starscream's principles generalize across sharding strategies and future architectures.

\end{abstract}

\section{Introduction}
\label{sec:introduction}
The computational demands of large language models (LLMs) have driven an unprecedented expansion in GPU capabilities. 
Modern accelerators such as AMD's Instinct MI300X offers a peak FP8 throughput of over  5~petaflops and 192~GB of HBM3 memory, while the following generation MI350 Series GPUs offers even greater performance through architectural refinements and superior memory bandwidth and capacity~\cite{mi350-roadmap}. This trajectory shows no signs of slowing: as foundation models grow from hundreds of billions to trillions of parameters, hardware vendors continue to scale accelerator designs to meet demand~\cite{gpu-scaling-trends}.

\begin{figure}[t]
\centering
\adjustbox{valign=c}{\includegraphics[width=0.42\columnwidth]{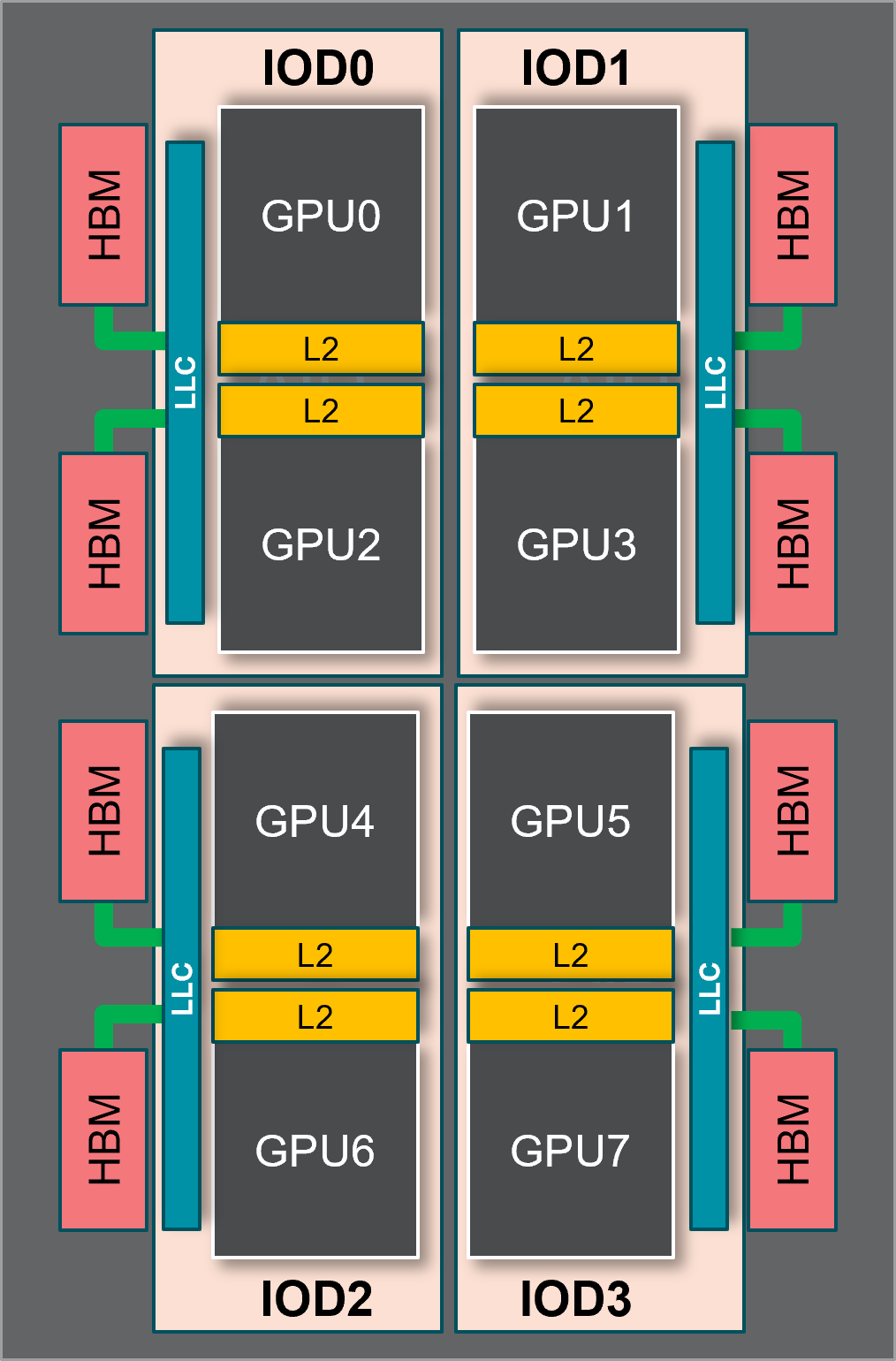}}
\hfill
\adjustbox{valign=c}{\includegraphics[width=0.56\columnwidth]{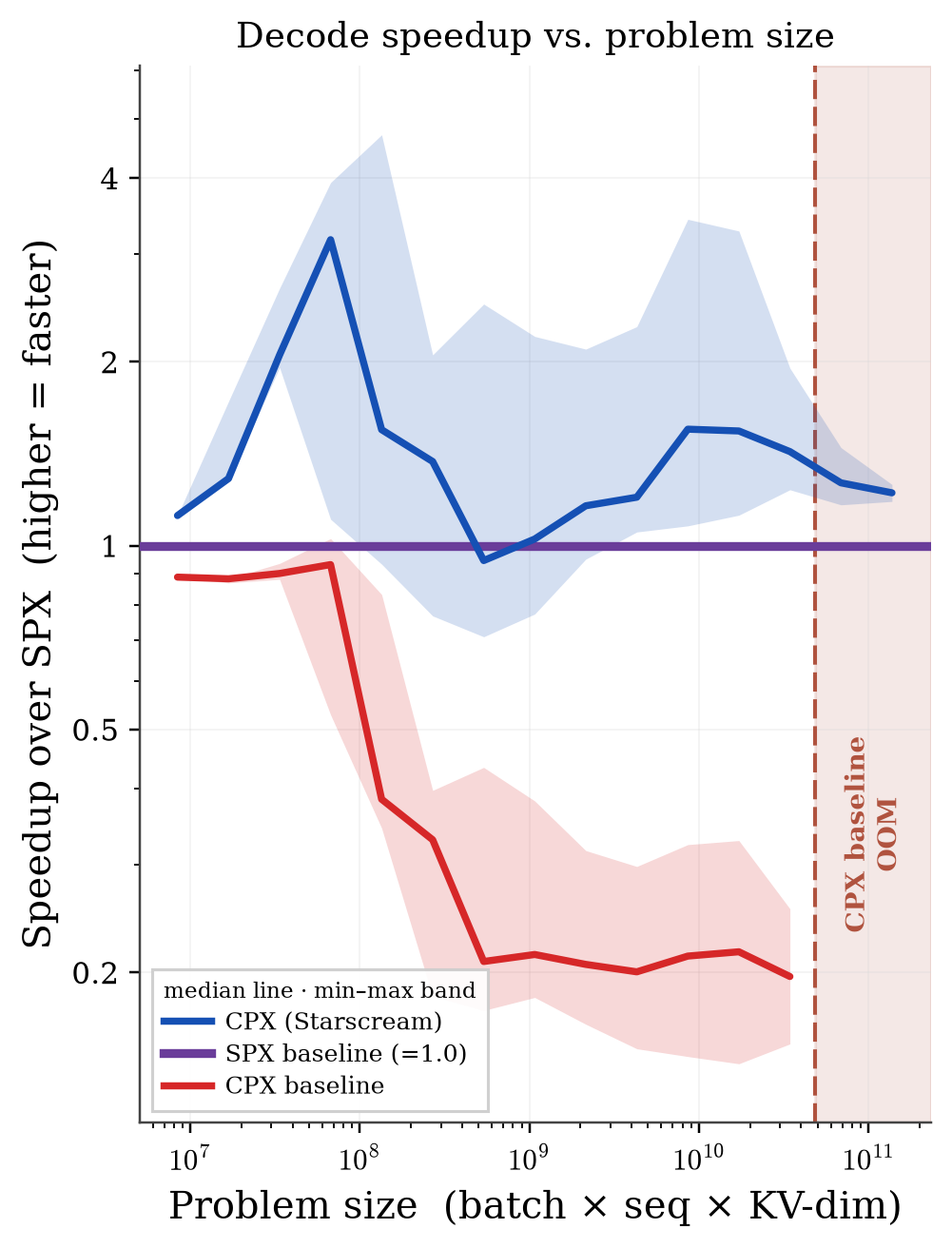}}

\hspace{2cm}
\scriptsize(A)
\hfill
\scriptsize(B)
\hspace{2cm}
\hfill

\vspace{-0.2cm}
\caption{\textbf{(A) AMD Instinct MI300X architecture. (B) Normalized self-attention time for Llama3-70B, Llama3-405B, and GPT-OSS120B on MI300X.
  Without partitioning: Purple line. With partitioning: red river is using naive sharding; blue river is using Starscream-enabled sharding. Bold colored lines show geomean
  speedup. }}
\label{fig:motivation}
\vspace{-0.5cm}
\end{figure}

However, manufacturing economics impose fundamental constraints on monolithic die sizes~\cite{interposer-micro16, mcm-gpu-isca17}. 
Larger dies suffer from exponentially decreasing yields as the probability of a defect 
appearing on any given die increases with area. Chiplet-based architectures have emerged 
as the industry's response to this challenge~\cite{chiplet-survey}. 
By decomposing a large monolithic design into multiple smaller dies interconnected through advanced packaging  technologies, vendors can maintain high yields while still delivering aggregate compute  density that would be hard with a single die. 
AMD's MI300X GPU exemplifies this 
approach: the MI300X integrates eight Accelerated Compute Dies (XCDs) on top of 4 I/O Dies (IODs) each connected to a subset of HBMs, all connected through a 
sophisticated on-package interconnect fabric. Future generations might expand this chiplet 
strategy further, potentially incorporating even more compute dies per package.

Alongside compute scaling, memory bandwidth has evolved in lockstep. HBM 
has progressed 
with each generation, delivering substantial 
improvements in both capacity and bandwidth~\cite{amd-cdna-roadmap}. The MI300X, for instance, 
provides up to 5.3~TB/s of aggregate HBM bandwidth, a figure that will only increase with subsequent generations (e.g., up to 8~TB/s in MI350 Series GPUs). 
Yet these impressive bandwidth figures mask a critical architectural reality: this bandwidth is not uniformly accessible from all
  compute units. As dies grow larger or adopt multi-chiplet designs, the physical distance between compute units and HBM
  controllers varies significantly—creating disparities between near and far memory access. This non-uniform memory access (NUMA)
  characteristic, long familiar in multi-socket CPU systems, is now becoming a first-class architectural concern in modern GPUs.

   AMD Instinct GPUs provide explicit mechanisms to exploit this inherent NUMA topology. Compute Partitioning modes
  logically divide the compute units (CUs) within a single physical GPU into independent virtual GPUs: DPX for
  2-way, QPX for 4-way, and CPX for per-XCD granularity (8-way on MI300X). Each virtual GPU is capable of
  running separate workloads with dedicated compute resources. Complementing compute partitioning, the Memory
  Partitioning (NPS) mode subdivides the HBM address space into NUMA domains. In NPS4 mode on MI300X, each NUMA
  domain encompasses the memory attached to a single IOD, enabling workloads to achieve higher effective bandwidth
  by localizing memory accesses to nearby controllers~\cite{amd-cpx-whitepaper}. We will refer to these capabilities collectively as \textit{partitioning}. Figure~\ref{fig:motivation}A illustrates this chiplet-based architecture.

  At peak isolation, these partitioning modes unlock significant architectural benefits. Operating MI300X in
  CPX+NPS4 mode (8 virtual GPUs mapped across 4 NUMA domains) with no cross-die
  communication, provides substantial advantages over the default SPX+NPS1 configuration~\cite{amd_compute_memory_modes, amd_mi300x_partitioning}:
  \begin{itemize}[leftmargin=*, nosep, topsep=0pt]
      \item \textbf{lower kernel launch latency:} Avoiding L2 cache flushes between kernel runs, formerly
   required to preserve coherence across XCDs.
      \item \textbf{over 20\% faster computations:} HBM accesses limited to the local die reduce signal travel distance
  and eliminate contention on shared resources~\cite{amd_compute_memory_modes, amd_mi300x_partitioning}.
      \item \textbf{Reduced IOD power consumption:} Lower cross-die traffic decreases power dissipation in the IODs. 
  \end{itemize}

  \noindent However, peak isolation is unrealistic for modern LLMs. Model parameters and KV caches exceed
  single-partition capacity, necessitating sharding across multiple devices, and how we shard determines what
  overheads get exposed. 
  
  Current frameworks—both inference
  (vLLM~\cite{vllm}, TensorRT-LLM~\cite{tensorrt-llm}) and training (Pytorch~\cite{fsdp,fsdp2},
  Megatron-LM~\cite{megatron-lm})—see virtual GPUs as independent devices, unaware of the partitioned topology.
  Total node resources remain unchanged, yet existing frameworks view virtual
  GPUs as additional devices. Naively extending the intra-node sharding strategies to the higher device
  count, though intuitive, exposes three fundamental bottlenecks: (1) reduced memory capacity from redundant memory replication, (2) degraded arithmetic intensity as matrices become increasingly narrow, and (3) communication
   overhead from flat collectives that ignore the bandwidth hierarchy between inter-die and inter-GPU links (inter-die links offer $10\times$
  inter-GPU bandwidth~\cite{amd-cpx-whitepaper, amd_rccl_usage}). 
  
  The reason for this drop is, established strategies deliver performance by sharding each tensor along a
  single dimension to minimize communication—tensor parallelism
  (TP) over weight matrices~\cite{megatron-lm}, sequence parallelism (SP) over inputs~\cite{sequence-parallel, context-parallel}, or FSDP for parameters and optimizer states~\cite{fsdp}. Beyond a certain parallelism degree, however, this unidimensional sharding
  exposes memory, compute, and communication bottlenecks, degrading performance.
These limitations manifest directly in partitioning configurations. 

  The net effect: partitioning exposes a new level of hierarchy that unlocks
  significant performance benefits. However, without such awareness, these advantages remain latent. Figure~\ref{fig:motivation}B (red-river)  shows an extreme example: 8 KV-head attention on 8 GPUs (64 in CPX), where the penalty is severe.

 We present Starscream, a rethinking of LLM execution for partitioned GPUs. 
\textit{The core insight:} leverage the coarse-grained control over data placement and compute mapping that partitioning modes provide to maximize data locality and minimize cross-die accesses. 
Rather than treating partitioned GPUs as simply ``more GPUs,'' Starscream recognizes that partitioning exposes a new level of hierarchy, and composes existing parallelism strategies hierarchically—applying each at the level of the topology where it is effective to extract performance from the same compute resources. 
This principle yields a systematic set of techniques:
 
  \begin{itemize}[leftmargin=*, nosep, topsep=0pt]
    \item \textbf{Eliminating memory pressure:} Hierarchical sharding avoids replicating data structures that naive sharding duplicates---KV caches in TP inference, parameters in FSDP training---preserving memory capacity and bandwidth.
    \item \textbf{High Arithmetic Intensity GEMMs:} 2D sharding maintains arithmetic intensity even as parallelism increases, keeping operations compute-bound rather than memory-bound.
      \item \textbf{Hierarchical collectives:} Collectives structured as two-level operations:
  fast intra-GPU operations over USR links, followed by inter-GPU synchronization on XGMI.
      \item \textbf{Computation-communication fusion:} Overlaps kernel execution (attention, GEMMs, 
      RMSNorm) with inter-XCD
      data movement, hiding collective latency behind computation.
  \end{itemize}

We implement and evaluate Starscream on AMD Instinct MI300X GPUs, demonstrating that
chiplet-aware design transforms GPU partitioning into a significant advantage.
Our NUMA-aware paged-attention kernels achieve up to \textbf{4$\times$
speedup} in CPX/NPS4 mode over SPX/NPS1. Our fused GEMM kernels with inter-XCD peer-mapping
deliver \textbf{1.43$\times$ faster execution}. These
per-kernel improvements compound to yield substantial end-to-end throughput gains across diverse
model architectures and batch sizes.  
Beyond raw performance, Starscream's hierarchical
sharding principles generalize across sharding strategies, offering a blueprint for exploiting
NUMA topologies in future chiplet-based accelerators.

The remainder of this paper is organized as follows. Section~\ref{sec:background} explains how partitioning yields performance benefits. Section~\ref{sec:motivation} presents the challenges that current LLM frameworks face under partitioning, identifying the root causes for performance degradation. Section~\ref{sec:starscream} describes the proposed Starscream hierarchical sharding strategies. Section~\ref{sec:results} details our experimental methodology and evaluation results demonstrating the effectiveness of our approach. Section~\ref{sec:related} shares the related works and we conclude in Section~\ref{sec:conclusion}.

\section{Partitioning on AMD Instinct MI300X GPUs}
\label{sec:background}

This section describes the AMD MI300X architecture and its compute/memory partitioning modes. While we focus on
  MI300X, the principles apply to most Instinct GPUs broadly.

\subsection{MI300X Chiplet Architecture}

Figure~\ref{fig:motivation}A illustrates the MI300X architecture.  At its core are four I/O Dies (IODs) connected via a ring-style Ultra Short Reach (USR) interconnect. The USR interconnect bandwidth is significantly higher than the XGMI bandwidth~\cite{amd_cpx_rccl_tuner, amd_rccl_usage}.

Each IOD hosts two Accelerator Compute Dies (XCDs) and connects to HBM memory slices. Within each XCD, 38 Compute Units (CUs) execute SIMD operations independently. Each XCD has a dedicated L2 cache slice, and a shared Infinity Cache spans each IOD, acting as the last-level cache (LLC).

\subsection{Baseline Operation: SPX NPS1}

In the default Single Partition Accelerator (SPX) mode with unified memory (NPS1), the GPU presents the illusion of a single monolithic device. The L2 cache appears shared across all XCDs, and HBM is accessible from any compute unit.

\textbf{Memory allocation.} Assuming a uniform memory distribution, pages would be allocated to HBM slices in round-robin fashion across all dies. The first page goes to HBM slice~0 on IOD~0, the second to HBM slice~1 on IOD~0, the third to HBM slice~0 on IOD~1, and so forth. 

\textbf{Data access paths.} When a CU requires data, the access latency depends on where the data resides. If the data is in a \textit{local} HBM slice---one connected to the same IOD as the requesting XCD---it is a direct read. However, if the data resides in a \textit{remote} HBM slice on a different IOD, the data must traverse the USR ring: it is first read from the remote HBM to its local IOD, then transferred across one or more USR hops to the requesting IOD, and finally delivered to the XCD. For the farthest memory (two hops away), this path is significantly longer than local access.

\subsection{Challenges in SPX NPS1 Mode}

Three challenges emerge from this unified memory model:

\noindent \textbf{Remote memory access overhead.} Since data is uniformly distributed across all HBM slices, a significant fraction of memory
   accesses target remote IODs, incurring the multi-hop latency described above. The problem compounds under concurrent access. When multiple XCDs simultaneously request data from different remote IODs, their traffic converges on the shared USR fabric, creating contention. The ring topology means that traffic between distant IODs must traverse intermediate links, competing with other flows. Even though HBM offers substantial aggregate bandwidth, the USR network becomes the bottleneck---effective bandwidth is limited by the slowest, most congested link. Combined, these two factors---longer data paths and USR contention---can account for over 20\% of total compute time.

\noindent \textbf{L2 cache coherence overhead.} Because L2 is sliced across XCDs, coherence must be maintained across dies. After a kernel completes, one XCD may hold dirty cache lines that remote XCDs are unaware of. Maintaining coherence requires flushing the L2 cache before each kernel launch, increasing launch latency by a non-trivial factor on MI300X.

\noindent \textbf{IOD power consumption.} Every remote HBM access requires the data to first travel from HBM through the IOD before reaching the USR interconnect. The IODs are thus actively involved in all cross-die data movement. Higher remote traffic translates to higher IOD power consumption, leaving less power budget for XCD compute operations. The GPU operates within a fixed power envelope, and power consumed by the IODs is power unavailable to the XCDs. With a reduced power budget, XCDs can only clock so high before hitting their power limit---resulting in lower operating frequencies and corresponding performance degradation.

\subsection{Compute and Memory Partitioning}

AMD Instinct GPUs can partition a single physical GPU into multiple compute and memory domains. We focus on the most aggressive configuration: CPX NPS4.

\textbf{Compute Partitioning (CPX)} mode partitions compute resources at XCD granularity. Each XCD becomes an
  independent virtual GPU, receiving 1/8th of total HBM capacity. The XCD's local L2 slice now becomes its private L2 cache.

\textbf{Memory Partitioning (NPS4).} NPS4 mode partitions memory into four NUMA domains, one per IOD. Each IOD's HBM slices constitute a separate NUMA domain. With CPX NPS4, each XCD's 1/8th memory allocation is exclusive to its local NUMA domain---the HBM slices connected to its IOD. Since each IOD hosts two XCDs, half of the IOD's HBM capacity is allocated to each XCD.

\textbf{Bandwidth implications.} In NPS1, each XCD can theoretically tap the aggregate bandwidth of all HBM slices across all IODs. In NPS4 on MI300X, each XCD can only access bandwidth from the HBM slices on the same IOD. This represents a trade-off: reduced aggregate bandwidth accessibility in exchange for guaranteed local bandwidth without contention.

\subsection{Benefits of CPX NPS4 Mode}

Operating in CPX NPS4 mode in perfect peak isolation eliminates the three challenges identified above:

\textbf{No L2 coherence overhead.} Each XCD has its own private L2 cache and operates as an independent GPU. There is no cross-die L2 sharing, no coherence concerns, and consequently no need for L2 flushes between kernel launches.

\textbf{No remote memory traffic.} In NPS4 mode, all memory requests from an XCD are routed exclusively to HBM slices within the same IOD. There is no cross-die memory traffic, eliminating USR contention and remote access latency.

\textbf{Reduced IOD power consumption.} Without cross-die data movement, the IODs consume less power, preserving power budget for XCD compute operations and enabling higher sustained frequencies.

These benefits motivate our focus on CPX NPS4 as the operating mode for GPU partitioning. However, as discussed in Section~\ref{sec:introduction}, realizing these benefits requires rethinking parallelization strategies---a challenge we address with Starscream.

\section{Anatomy of Na\"{\i}ve Sharding with Partitioning}
\label{sec:motivation}

We motivate the challenges of partitioning through a concrete example: deploying a Grouped Query Attention (GQA)~\cite{gqa} model using tensor parallelism~\cite{megatron-lm} on an illustrative node with GPUs supporting compute and memory partitioning. This example exposes the three fundamental bottlenecks discussed earlier: data replication, arithmetic intensity degradation, and collective communication overhead; that cause naive sharding to degrade performance despite its architectural advantages.

\subsection{Illustrative System Setup and Model Characteristics}

\begin{figure}[hbt]
  \centering
    \adjustbox{valign=c}{\includegraphics[width=0.9\columnwidth]{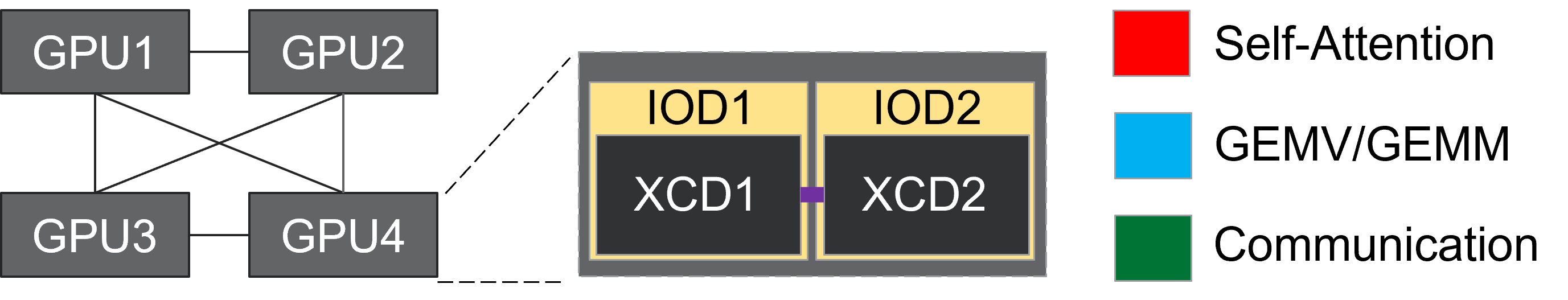}}
    
    \small(A)~Illustrative node
    
  \vspace{0.1cm}
  \vspace{0.1cm}
      \adjustbox{valign=c}{\includegraphics[width=\columnwidth]{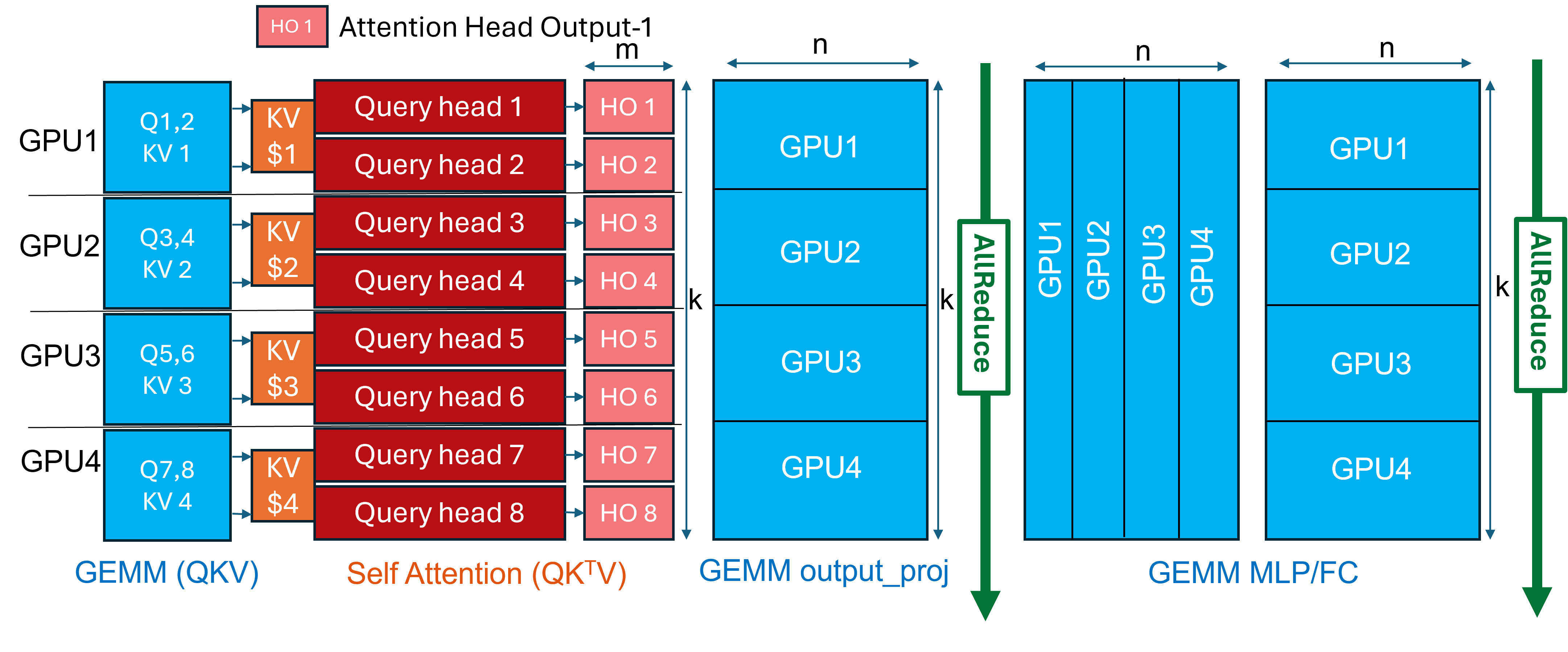}}
(B)~Illustrative LLM model
\vspace{-0.2cm}
\caption{\textbf{Illustrative (a) partitioned node and (b) LLM model.}}
\label{fig:illustrative model}
  \vspace{-0.6cm}
  \end{figure}

\noindent  \textbf{Hardware Configuration.} We consider a node equipped with 4 physical GPUs interconnected via AMD's
  Infinity Fabric (XGMI)~\cite{amd-infinity-fabric}, shown in Figure~\ref{fig:illustrative model}A. Each GPU
  contains 2 IODs and 2 XCDs (one per IOD), linked by high-bandwidth USR connections. In the baseline SPX/NPS1 mode,
   software frameworks perceive 4 monolithic GPUs. Enabling DPX/NPS2 mode partitions each physical GPU into 2 virtual GPUs, exposing 8 logical devices to the framework,
  each with half the memory capacity and bandwidth.

\noindent \textbf{Model Configuration.} Our illustrative LLM layer uses Grouped-Query Attention (GQA) with 8 attention heads
   and 4 KV heads, where each KV head is shared by 2 query heads. The layer's computation comprises three kernel
  categories: \textit{(1) Attention kernels} perform self-attention by querying the KV cache; \textit{(2) GEMM
  kernels} handle input projection (QKV generation), output projection, and two MLP layers (FC1, FC2); \textit{(3)
  Collective kernels} synchronize partial results across GPUs via all-reduce operations. While modern LLMs include
  additional kernel types, we focus on those that dominate execution time in this illustration.
  Figure~\ref{fig:illustrative model}B illustrates these components and their tensor-parallel sharding under the
  baseline 4-GPU configuration.

\vspace{-0.2cm}

\begin{figure*}[t]
\centering
\vspace{-0.1cm}
\adjustbox{valign=c}{\includegraphics[scale=0.36]{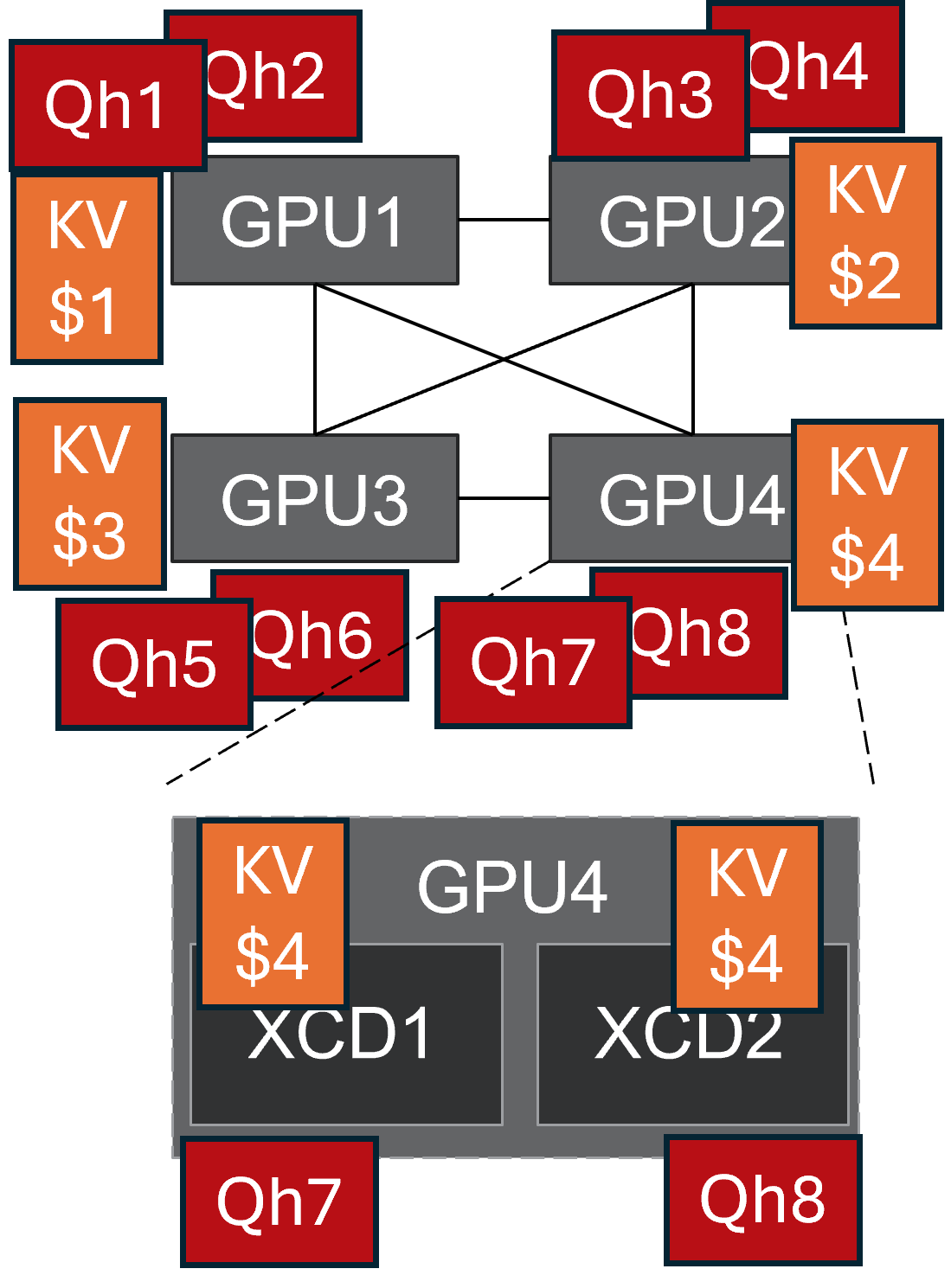}}
\hfill
\vrule
\hfill
\adjustbox{valign=c}{\includegraphics[width=0.9\columnwidth]{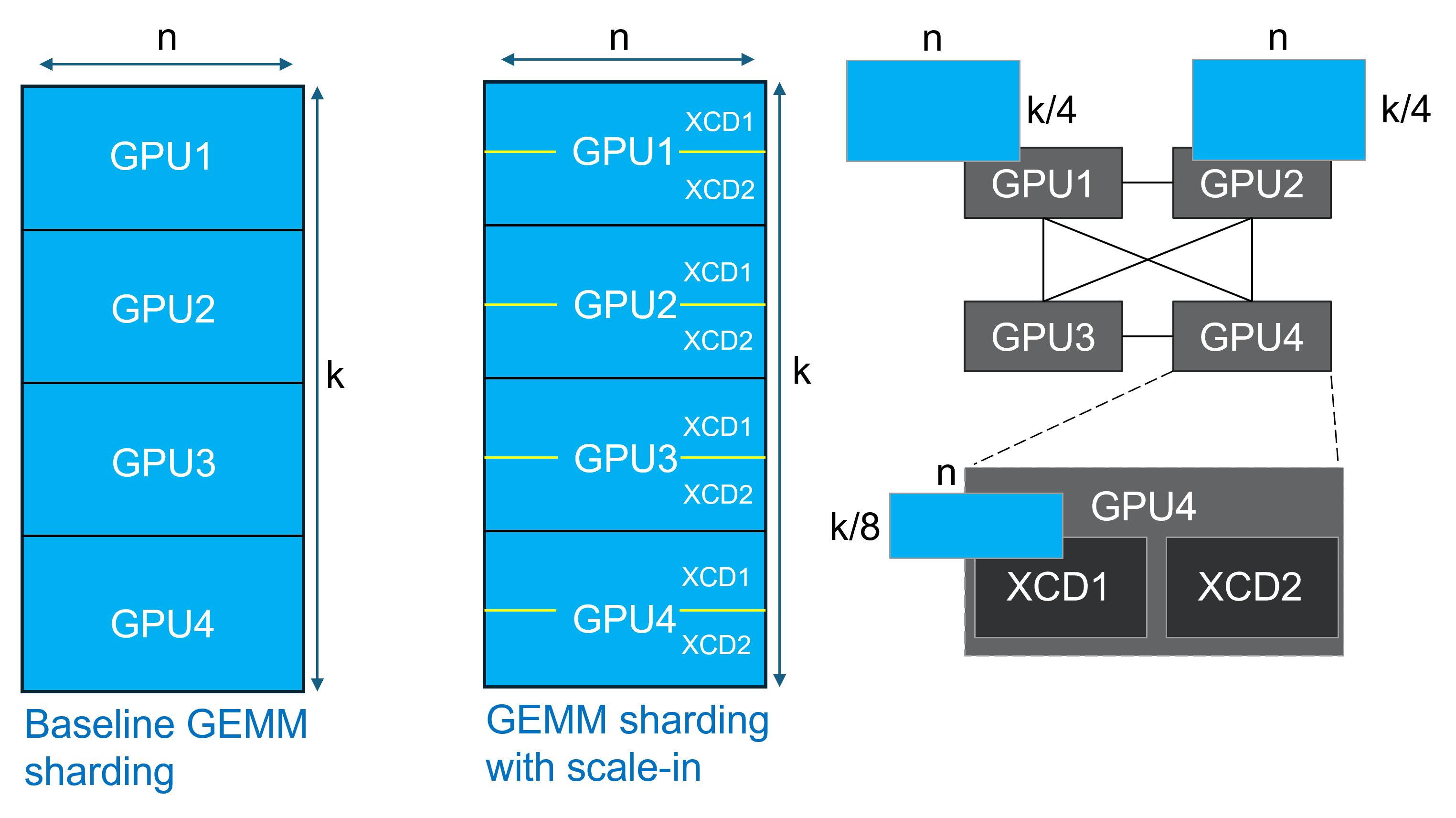}}
\hfill
\vrule
\hfill
\adjustbox{valign=c}{\includegraphics[scale=0.34]{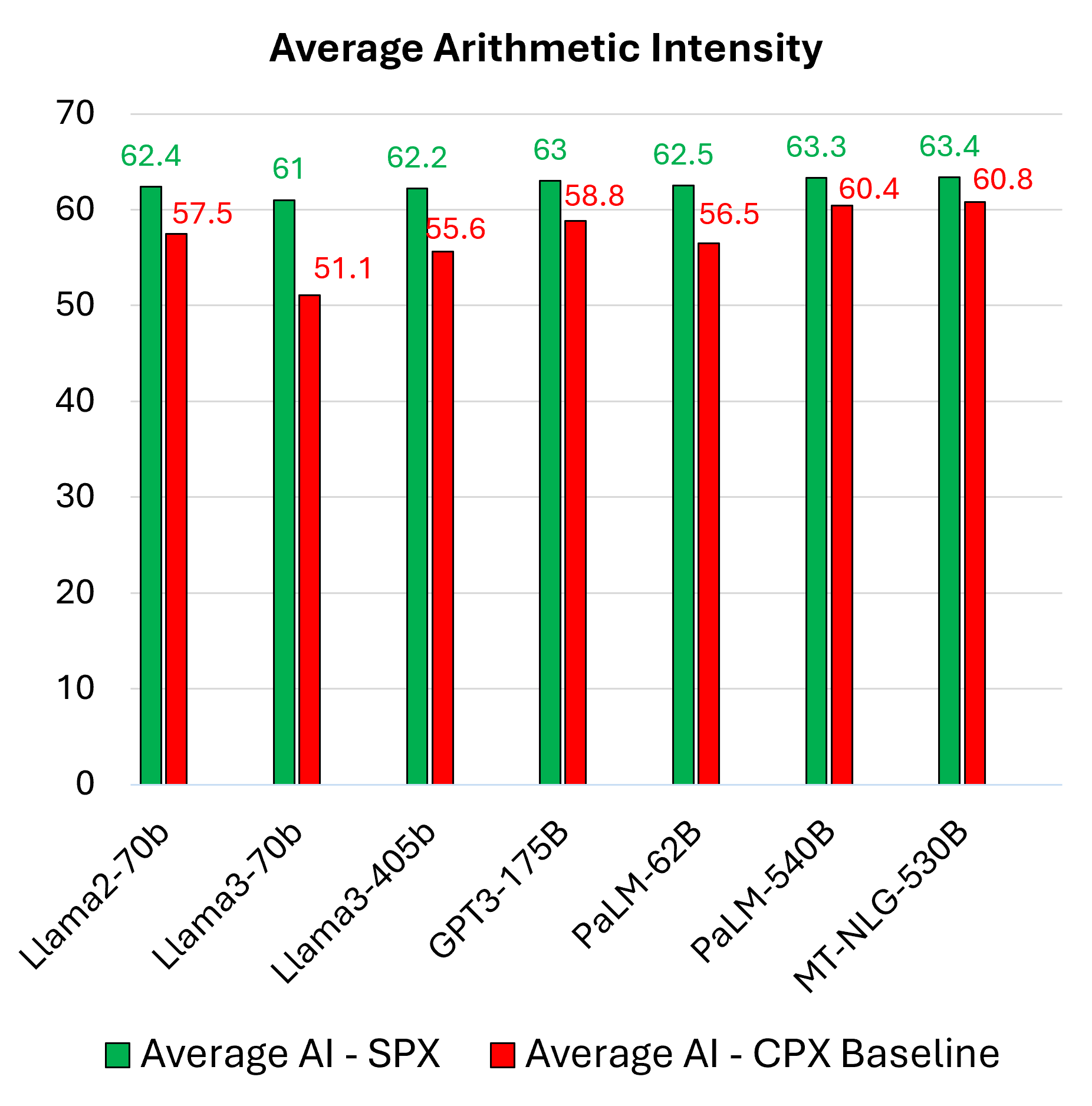}}

\hspace{2cm}
\small(A)
\hfill
(B)
\hfill
(C)
\hspace{2cm}
\hfill

\vspace{-0.2cm}
\caption{\textbf{Illustrative GPU sharding for (A) GQA and (B) RowParallelLinear. (C) Arithmetic intensity: 8 GPUs operating in SPX versus CPX.}}
\label{fig:motivation-detail}
\vspace{-0.5cm}
\end{figure*}

\subsection{Attention Kernels: KV Cache Replication}

   Tensor parallelism distributes attention computation via \textit{head-parallel sharding}: each GPU receives a
  subset of query heads along with their corresponding KV heads. In the baseline 4-GPU configuration
(Figure~\ref{fig:illustrative model}A), each GPU handles 2 query heads and 1 KV head (top half of
Figure~\ref{fig:motivation-detail}A). Since both query heads share the same KV head for their self-attention
  computation, and all three reside on the same GPU, this sharding requires no cross-GPU communication and
  incurs no redundant memory overhead.

Partitioning disrupts this clean mapping. With 8 virtual GPUs, head-parallel sharding assigns only 1 query head per
  device, but only 4 KV heads exist to serve 8 devices. If we naively increase tensor parallelism from 4 to 8 to
  match the new device count, current frameworks (vLLM~\cite{vllm}, Megatron-LM~\cite{megatron-lm}) resolve this
  mismatch by \textit{replicating} KV heads: each virtual GPU receives a full copy of its required KV head (bottom half of Figure~\ref{fig:motivation-detail}A). Consequently, each physical GPU now stores the same query heads as
  before (1 per XCD), but maintains \textbf{2 redundant copies} of the same KV head across its two XCDs.

This replication carries severe consequences. The KV cache already consumes 3$\times$ more memory than model
  parameters in long-context scenarios~\cite{pope2022efficientlyscalingtransformerinference}. Duplicating it across
  virtual GPUs \textbf{reduces effective memory capacity} per physical GPU, directly limiting maximum batch size and
  supported context length. Worse, both XCDs now read from separate local copies rather than sharing cached data,
  \textbf{halving effective memory bandwidth}. On a real MI300X with 8-way partitioning, this replication amplifies
  to 8$\times$, significantly reducing both available memory and bandwidth. Since attention is memory-bandwidth
  bound during autoregressive decode, execution time increases proportionally---and the system hits out-of-memory
  far sooner than baseline, as shown in Figure~\ref{fig:motivation}B.

\subsection{GEMM Kernels: Arithmetic Intensity Collapse}

  Tensor parallelism~\cite{megatron-lm} shards GEMM operations using two complementary strategies.\footnote{Terms \textit{ColumnParallelLinear},  \textit{RowParallelLinear} used from vLLM~\cite{vllm}.} Consider a
  GEMM computing $C_{M \times N} = A_{M \times K} \cdot W_{K \times N}$ distributed across $P$ GPUs:

  \begin{itemize}[leftmargin=*, nosep, topsep=0pt]
      \item \textbf{ColumnParallelLinear} shards the weight matrix along the column dimension (i.e., $N$). Each
  GPU stores $W_{K \times N/P}$ and receives a full copy of the input $A_{M \times K}$, then independently computes
  a non-overlapping output slice $C_{M \times N/P}$. FC1, the first MLP layer, uses this strategy (Figure~\ref{fig:illustrative model}B).

      \item \textbf{RowParallelLinear} shards along the row dimension (i.e., $K$). Each GPU receives input $A_{M
   \times K/P}$---matching the sharded output from FC1---and stores a correspondingly sharded weight $W_{K/P \times
  N}$. Each GPU computes a full $C_{M \times N}$ result, but because $K$ is partial, these must be aggregated via
  all-reduce. This strategy is typically used in the second MLP layer (FC2) in Figure~\ref{fig:illustrative model}B.
  \end{itemize}

  We focus on RowParallelLinear, illustrated in the last layer of Figure~\ref{fig:illustrative model}B. In the baseline 4-GPU
  configuration, each GPU receives a $K/4 \times N$ weight shard. Under partitioning with 8 virtual GPUs
  (Figure~\ref{fig:motivation-detail}B), each receives $K/8 \times N$. On a typical MI300X node with 8
  physical GPUs operating in CPX mode, this degrades further to $K/64 \times N$ per virtual GPU.

  The problem is that these sharding strategies employ \textit{unidimensional sharding}: sharding typically occurs
  along the same dimension ($K$ for RowParallelLinear, $N$ for ColumnParallelLinear) regardless of parallelism
  degree. As $P$ grows, this produces increasingly \textbf{skinny, elongated matrices} that are inefficient to
  compute. A $K/64 \times N$ shard has dramatically lower arithmetic intensity than the original $K \times N$
  matrix, even though total computation and hardware resources remain constant. 
  Operations compute-bound at $P=1$ might become memory-bound at $P=64$, underutilizing available FLOPS.
  Figure~\ref{fig:motivation-detail}C quantifies this drop for 8-way versus 64-way sharding across various
  models.

\subsection{Collective Kernels: Quadratic Communication Growth}

RowParallelLinear requires all-reduce to aggregate partial outputs across GPUs. In the baseline 4-GPU
  configuration (Figure~\ref{fig:illustrative model}B), each GPU holds an $M \times N$ partial result and
  exchanges data with 3 peers. Assuming ring all-reduce, each GPU sends and receives $(P-1) \cdot M \cdot N$ elements.

  Partitioning dramatically increases the peer count. With 8 virtual GPUs, each must now communicate with 7 peers,
  sending $(8-1) \cdot M \cdot N$ elements. Although total computational work remains constant,
  \textbf{communication volume grows with parallelism degree}: doubling $P$ from 4 to 8 increases per-GPU
  data movement from $3 \cdot M \cdot N$ to $7 \cdot M \cdot N$---a $2.3\times$ increase for a $2\times$ increase in
   parallelism. For 64-way parallelism on an 8-GPU MI300X node, each virtual GPU exchanges data with 63 peers,
  resulting in $21\times$ more communication than the baseline. While more advanced all-reduce algorithms can
  mitigate this overhead, the fundamental scaling problem persists.
  Moreover, naive sharding treats all virtual GPUs uniformly, ignoring the heterogeneous link topology.
  Communication between virtual GPUs on the same physical die (via USR links) enjoys $10\times$ higher bandwidth
  than cross-die communication (via XGMI)~\cite{amd-cpx-whitepaper}. Current frameworks perform flat all-reduces
  that fail to exploit this hierarchy: all transfers proceed at the rate of the slowest link (XGMI), leaving the
  faster intra-die bandwidth (USR) underutilized.
  
\subsection{Training: FSDP2 Exhibits Identical Pathologies}

While the preceding analysis focused on tensor-parallel inference, training workloads using Fully Sharded Data Parallel (FSDP)~\cite{fsdp,fsdp2} exhibit identical pathologies. FSDP shards model parameters, gradients, and optimizer states across GPUs while operating on a different batch. During forward and backward passes, each GPU all-gathers layer parameters before computation and
  reduce-scatters gradients after, respectively.

In the baseline 4-GPU configuration, frameworks overlap communication with computation: while GPU $i$ computes layer $\ell$, it concurrently all-gathers parameters for layer $\ell+1$. Similarly, during backward passes, reduce-scatter for layer $\ell$ overlaps with gradient computation for layer $\ell-1$. This pipelining masks communication latency, though some exposure remains for large models or small micro-batches.

Partitioning breaks this balance. Each virtual GPU now all-gathers the \textit{entire} layer's parameters to its local memory---which has \textbf{half the capacity and bandwidth} of the original physical GPU. For an 8-way partitioned MI300X, this means each XCD must reconstruct full layer parameters into 1/8th the memory with 1/8th the bandwidth. The collective scope also increases from 3 peers (4-GPU baseline) to 7 peers (8-way sharding), amplifying communication overhead.

The arithmetic intensity problem also persists. As $P$ grows, the data batch per $P$ become increasingly skinny, reducing per-GPU arithmetic intensity and underutilizing compute resources during local forward/backward computations.

  \begin{tcolorbox}[title={\textbf{Summary: Universal Bottlenecks}}, colback=gray!6, colframe=gray!100,
  fonttitle=\bfseries]
  The bottlenecks identified---memory capacity reduction, collective communication overhead, and arithmetic
  intensity degradation---are \textit{universal across sharding strategies and workload types}. Whether applying
  tensor parallelism or FSDP2, whether performing inference (prefill or decode) or training (forward or backward),
  naive sharding on partitioned GPUs amplifies the same fundamental issues:

  \begin{itemize}[leftmargin=*, nosep, topsep=0pt]
      \item \textbf{Memory pressure:} Fractional capacity/bandwidth limits batch size, context length, and model scale.
      \item \textbf{Communication inflation:} Data movement scales with partitioning; flat collectives skip link
  hierarchy.
      \item \textbf{Compute underutilization:} Unidimensional sharding yeilds skinny, low arithmetic
  intensity matrices.
  \end{itemize}


  These pathologies compound to overwhelm partitioning's architectural benefits (significantly faster kernel launch,
  low latency accesses, and higher power window), causing significant end-to-end performance degradation compared to the SPX/NPS1 baseline.
  \end{tcolorbox}

\section{Starscream}
\label{sec:starscream}

Partitioning fundamentally transforms the hardware substrate. A single physical GPU becomes several smaller virtual GPUs---each with reduced memory capacity, diminished bandwidth, a fraction of the L2 cache, and fewer compute units. These are not minor perturbations; they reshape the performance characteristics that every sharding strategy was tuned to exploit.

Every parallelism strategy carries inherent overheads that amplify or remain hidden depending on hardware characteristics. When partitioning changes these characteristics, bottlenecks emerge---affecting enough kernels to degrade end-to-end execution. Section~\ref{sec:motivation} quantified these pathologies.                          

Starscream addresses this gap. By systematically analyzing the architectural changes that partitioning introduces, Starscream defines fundamental principles that capture the benefits of partitioning while paying minimal cost. The key insight is deceptively simple: \textit{do not extend existing sharding deeper---shard orthogonally instead}.                                                   

We present Starscream in two parts. First, we introduce the \textbf{orthogonal sharding principle}---the theoretical foundation that explains why naive sharding fails and what alternative succeeds. Second, we apply this principle to \textbf{tensor-parallel inference}, detailing how orthogonal sharding transforms self-attention, GEMMs, and collectives on AMD hardware.

\subsection{The Orthogonal Sharding Principle}

The key insight behind Starscream is \textit{multi-dimensional sharding}. When workloads are sharded across physical GPUs in a node along a given dimension, switching to partitioning should \textbf{not} always extend that same sharding deeper---this is precisely what causes naive sharding to fail. Instead, partitioning should retain the work per physical GPU exactly as the ideal sharding for that GPU topology suggests, while introducing sharding along an \textit{orthogonal dimension} across the virtual GPUs within each physical GPU.

\begin{figure}[h]
\centering
\includegraphics[width=0.8\columnwidth]{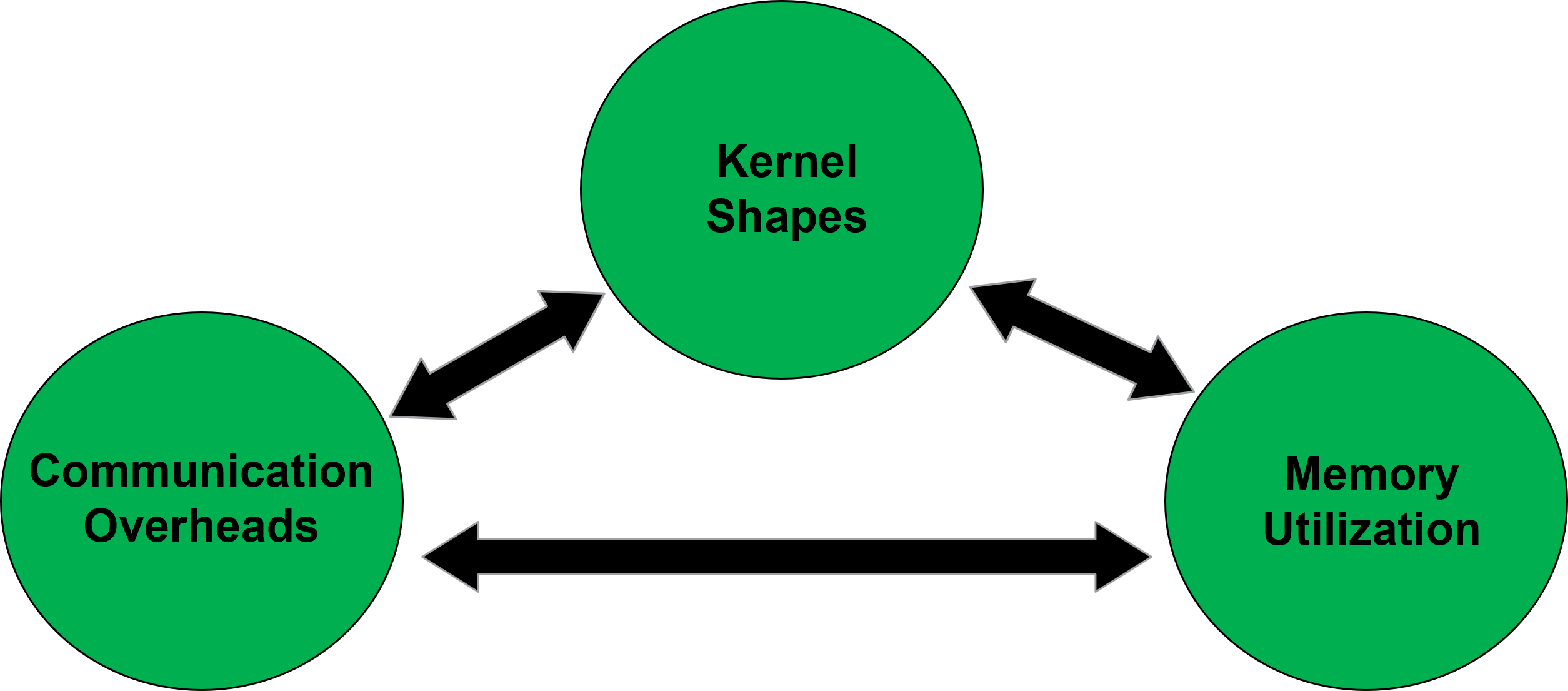}
\caption{\textbf{The three tradeoffs guiding orthogonal sharding.}}
\label{fig:starscream-overview}
\end{figure}

This orthogonal dimension is chosen to simultaneously optimize three critical tradeoff points, as illustrated in Figure~\ref{fig:starscream-overview}:

\begin{enumerate}
    \item \textbf{Kernel shapes:} The orthogonal shard must produce matrix tiles that maintain high arithmetic intensity, avoiding the skinny elongated shapes that shift operations from compute-bound to memory-bound.

    \item \textbf{Memory redundancy:} The orthogonal shard must avoid replicating tensors (such as KV caches) that would consume precious memory capacity and bandwidth.
    \item \textbf{Communication overhead:} The orthogonal shard must minimize cross-die data movement time by exploiting the hierarchical bandwidth structure---fast intra-GPU links versus slower inter-GPU links.
\end{enumerate}

The choice of \textit{which} orthogonal dimension to shard along under partitioning is determined by analyzing these three tradeoff points for the specific inter-GPU sharding strategy in use. Different parallelism strategies---tensor parallelism (TP), fully sharded data parallelism (FSDP), sequence parallelism (SP)---each shard along different dimensions across GPUs, and consequently each requires a different orthogonal dimension for effective performance. 

Crucially, orthogonal sharding is \textit{compositional}: 
Starscream does not replace existing parallelism strategies---it augments them. A framework can continue using tensor parallelism, FSDP, or any combination across physical GPUs exactly as before; Starscream adds NUMA-aware partitioning at the intra-GPU level. This design philosophy transforms GPU partitioning into an opportunity to extract additional performance from the same compute resources.

\subsection{Scope: Tensor Parallelism as a Case Study}

While Starscream provides a comprehensive framework with solutions for each sharding strategy (TP, FSDP, SP, PP, and their combinations), the underlying principles are universal. Given the limited space in this paper, we focus on \textbf{tensor parallelism for LLM inference} as a detailed case study. This choice is deliberate:

\begin{itemize}
    \item Tensor parallelism is the dominant strategy for large-model inference, where latency-sensitive decode phases make partitioning's locality benefits most attractive.

    \item The three bottlenecks---memory replication, GEMM intensity collapse, and collective overhead---manifest most acutely under tensor parallelism with Grouped Query Attention models~\cite{gqa}.

    \item The orthogonal sharding solutions for tensor parallelism directly demonstrate how analyzing the three tradeoff points guides dimension selection.
\end{itemize}

Since the principles are fundamental, the insights from this case study transfer directly to other sharding strategies and attention variants ( MQA~\cite{mqa}, MLA~\cite{mla-deepseek}, CHAI~\cite{chai}). For example, extending FSDP to partitioning scope would require all-gathering parameters into smaller memory while increasing microbatch counts, both compound to burden the memory capacity limiting the max model size supported. The orthogonal approach instead shards along dimensions (hidden, sequence) that avoid expanding all-gather scope. The analysis framework remains identical; only the specific dimension choices differ.

\subsection{Self-Attention with Sequence-Length Sharding}
\label{subsec:attention}

As established in Section~\ref{sec:motivation}, tensor-parallel self-attention results in KV cache replication when the number of KV heads is fewer than the number of virtual GPUs---a common scenario with 64 virtual GPUs per node in CPX/NPS4 mode and modern architectures using GQA or MLA. Since memory capacity does not scale with partitioning, naive sharding creates multiple redundant copies of the same KV cache, degrading both capacity and bandwidth.

\begin{figure}[t]
\centering
\includegraphics[width=\columnwidth]{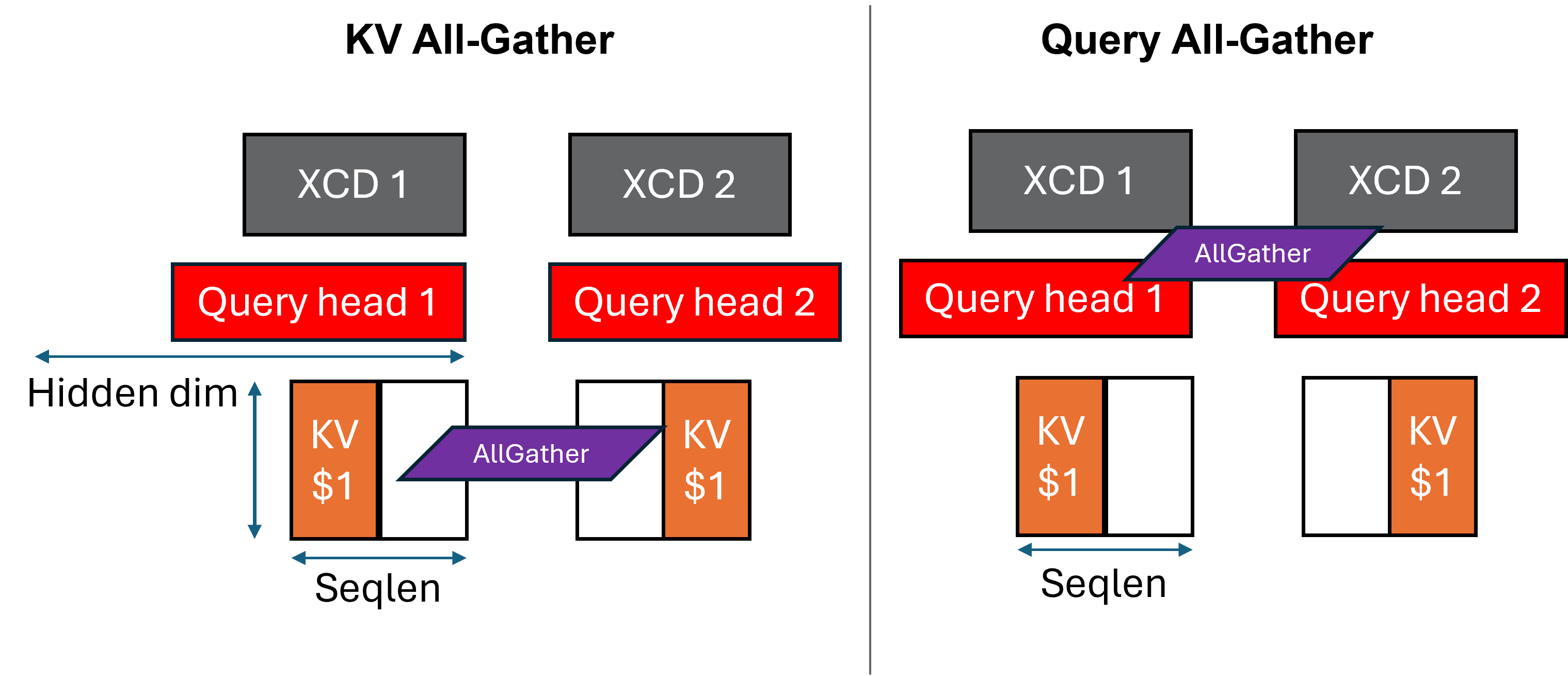}
\vspace{-0.4cm}
\caption{\textbf{Starscream sequence-length sharding: two self-attention variants.}} 
\vspace{-0.5cm}
\label{fig:attention-sharding}
\end{figure}

\paragraph{Orthogonal Sharding Strategy} Instead of replicating the KV cache, Starscream shards it across compute partitions along the \textit{sequence length} dimension. Each virtual GPU holds a subset of tokens, allocated in round-robin fashion: the first token's KV entry resides on XCD0, the second on XCD1, and so on. Figure~\ref{fig:attention-sharding} illustrates this sharding. Each XCD retains its assigned query head from inter-GPU tensor parallelism, but rather than duplicating the KV cache, XCDs share a single distributed copy. This eliminates memory redundancy entirely.

\paragraph{Computation with Distributed KV Cache} Since self-attention requires each query to attend over \textit{all} KV tokens, the distributed cache necessitates coordination. Two approaches are possible:

\begin{enumerate}
    \item \textbf{KV all-gather:} Before computing each layer, all-gather the KV cache entries across XCDs. Each XCD then holds a complete copy, computes full self-attention locally, and evicts the gathered entries afterward---analogous to how FSDP all-gathers parameters per layer.                                                                  
    \item \textbf{Query all-gather with partial attention:} Keep KV entries in place and all-gather the query heads instead. Each XCD then holds all query heads but only its local KV subset, computing \textit{partial} self-attention. These partials are then exchanged via a collective call, followed by a lightweight aggregation kernel.
\end{enumerate}

\begin{figure*}[t]
\centering
\adjustbox{valign=c}{\includegraphics[width=0.3\linewidth]{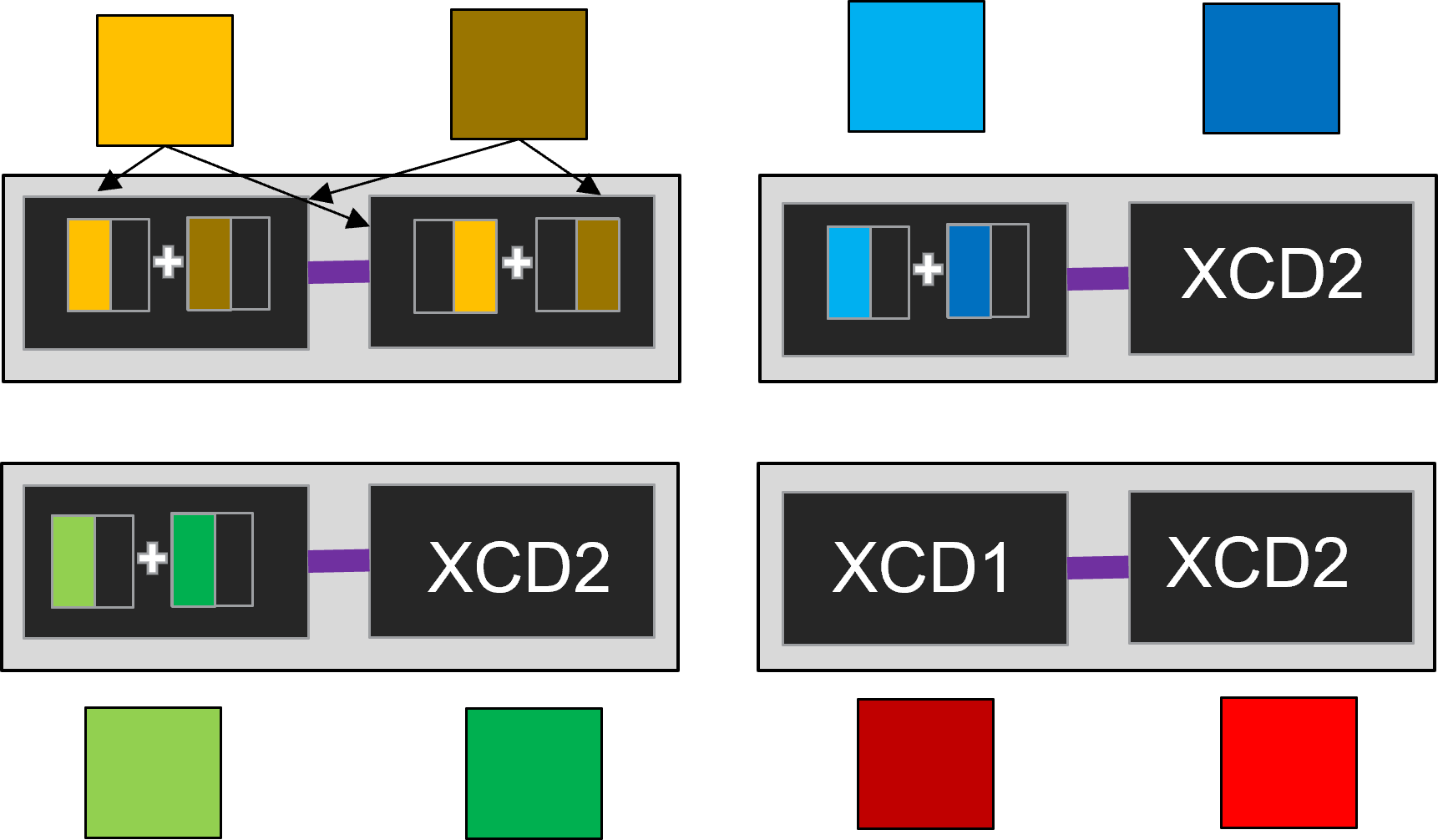}}
\hfill
$\Rightarrow$
\hfill
\adjustbox{valign=c}{\includegraphics[width=0.3\linewidth]{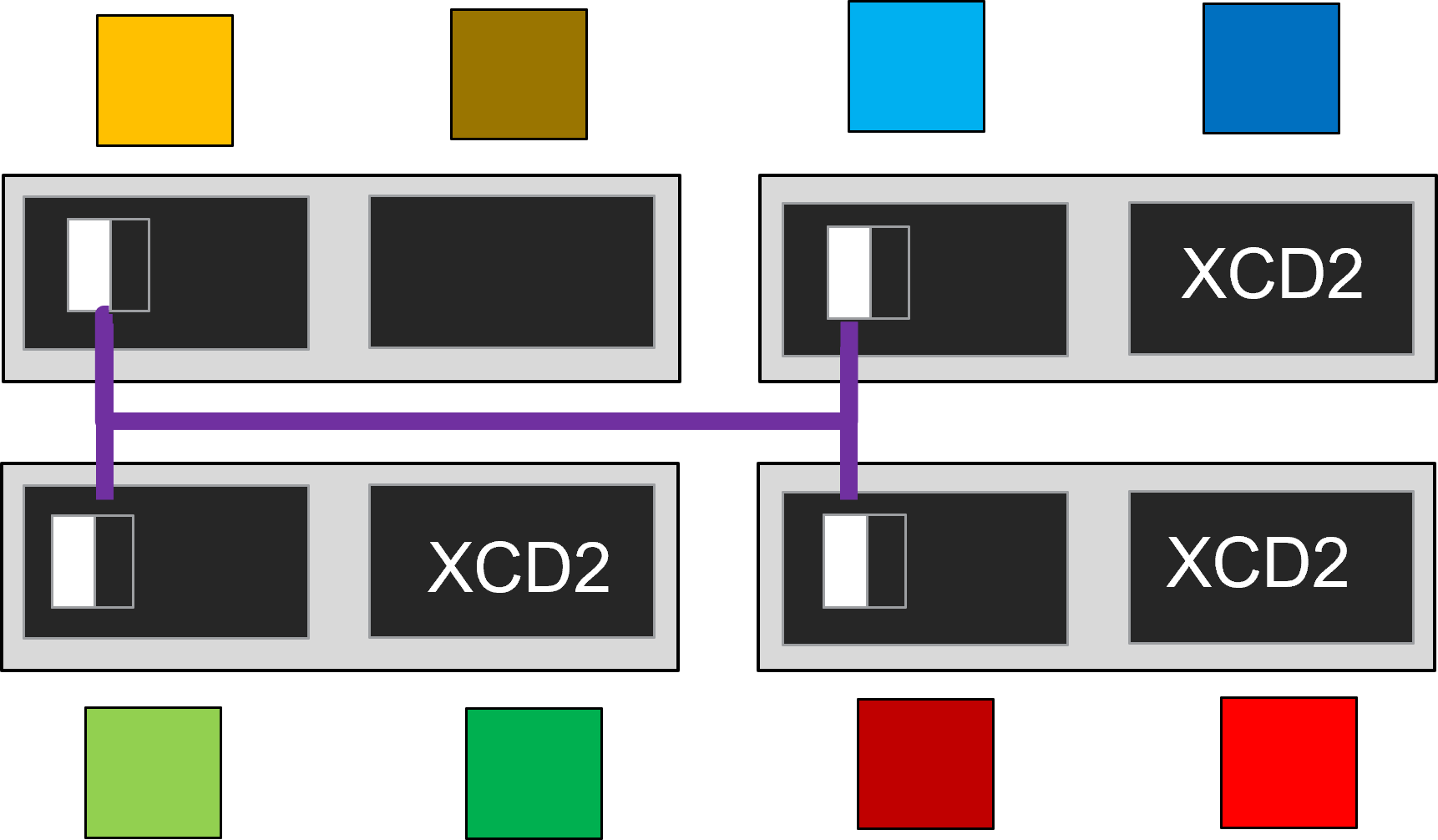}}
\hfill
$\Rightarrow$
\hfill
\adjustbox{valign=c}{\includegraphics[width=0.3\linewidth]{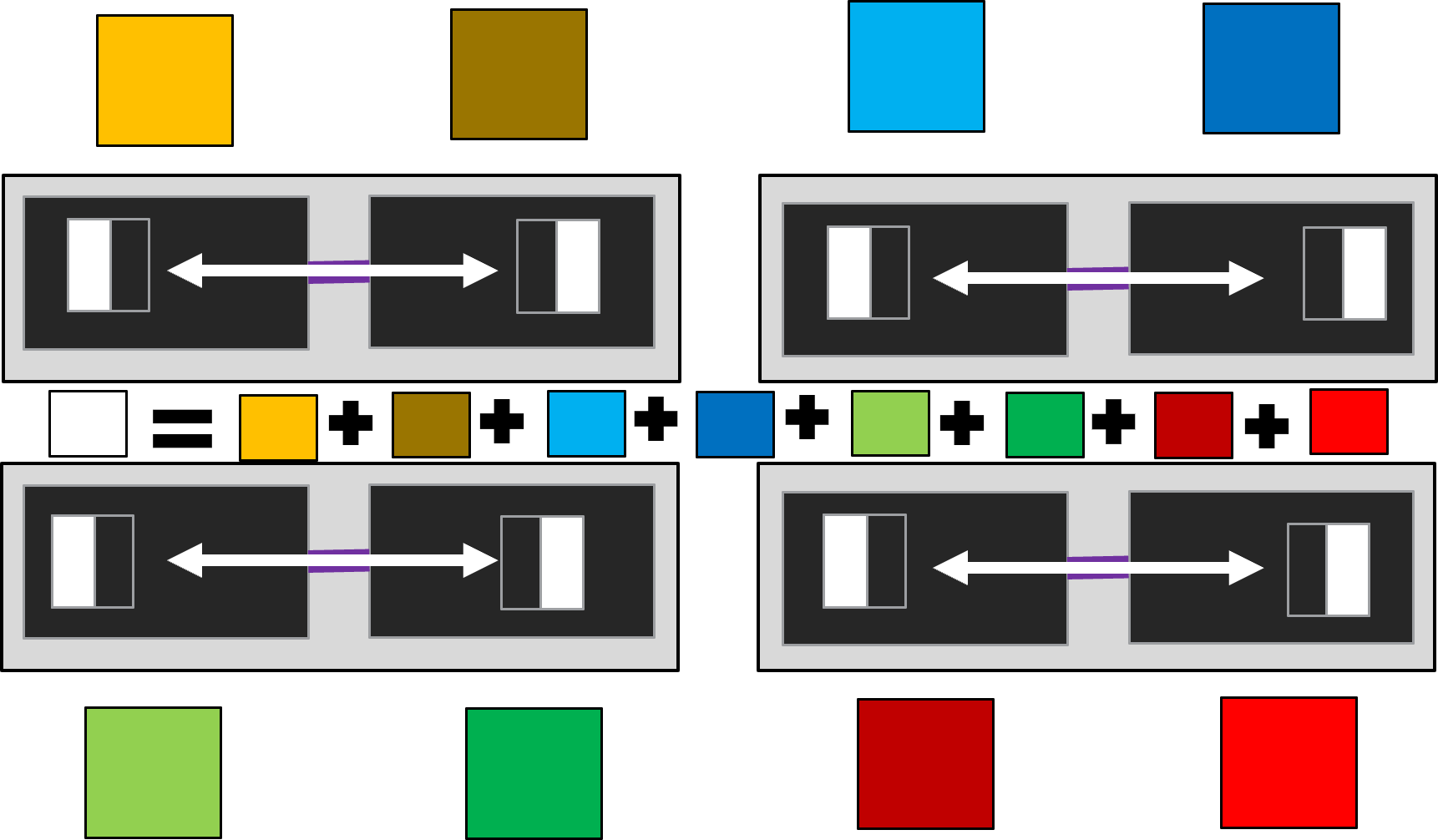}}

\hspace{2.5cm}
\small(A)
\hfill
(B)
\hfill
(C)
\hspace{2cm}
\hfill

\vspace{-0.2cm}
\caption{\textbf{The three-stage hierarchical all-reduce: (A) intra-GPU reduce-scatter, (B) inter-GPU peer-XCD all-reduce, (C) intra-GPU all-gather.}}
\label{fig:hierarchical-collectives}
\vspace{-0.5cm}
\end{figure*}

\paragraph{Decode versus Prefill} The choice between these approaches depends on the inference phase. During \textbf{decode}, the autoregressive nature produces exactly one query token per iteration, while the KV cache holds all tokens observed thus far. Moving queries is far cheaper than moving the entire KV cache---we select query all-gather. During \textbf{prefill}, all input tokens enter simultaneously, so query and KV token counts are comparable. However, GQA models have far more query heads than KV heads (e.g., Llama-3-405B: 128 query heads, 8 KV heads). Since KV head volume is 16$\times$ smaller, we select KV all-gather for prefill. Notably, KV all-gather enables full local attention, eliminating the aggregation kernel.

\paragraph{Effectiveness} This approach dramatically reduces inter-die data movement. Consider Llama-3-405B on MI300X in CPX/NPS4 (64 virtual GPUs) with 128K context. During decode, transferring only query heads and aggregation metadata accounts for less than 0.1\% of the data consumed by the attention computation. During prefill, KV all-gather accounts for only 4.4\% of total data---a 16$\times$ reduction from query all-gather due to the GQA head ratio. Compared to the baseline where 75\% of memory accesses were remote (Section~\ref{sec:background}), Starscream reduces cross-die traffic to under 5\%.

Beyond reduced latency, sequence-length sharding delivers two additional benefits. First, since each KV entry is read exactly once per physical GPU (on whichever XCD owns it), memory bandwidth is no longer wasted reading duplicate copies. Second, inter-XCD traffic occurs via controlled collective calls rather than scattered remote accesses, avoiding contention and maximizing HBM bandwidth utilization.

\subsection{Hierarchical Collective Primitives}
\label{subsec:collectives}

As discussed in Sections~\ref{sec:background} and~\ref{sec:motivation}, conventional collective calls treat all links uniformly, ignoring the bandwidth difference between much faster USR links and XGMI links. When naive sharding splits workloads across all XCDs in a node, partials must be aggregated across the entire topology. A flat all-reduce spanning all 64 virtual GPUs in MI300X bottlenecks on the slowest link---XGMI---leaving the faster USR fabric underutilized.

\paragraph{Three-Phase Hierarchical All-Reduce} Starscream restructures all-reduce as a three-phase operation that decouples intra-GPU and inter-GPU communication. We illustrate using the 4-GPU, 2-XCD-per-GPU system from Figure~\ref{fig:motivation-detail}, where each colored tile represents an $M \times N$ partial result.

\begin{enumerate}
    \item \textbf{Intra-GPU reduce-scatter (Phase 1):} Each physical GPU performs a local reduce-scatter across its XCDs over USR links. As shown in Figure~\ref{fig:hierarchical-collectives}A, after this phase each XCD holds half the tile (shape $M \times N/2$), containing the reduced result from both local XCDs. All physical GPUs execute this phase simultaneously with no cross-GPU dependency.

    \item \textbf{Inter-GPU peer-XCD all-reduce (Phase 2):} XCDs at the same position across physical GPUs (all XCD0s, all XCD1s) perform an all-reduce over XGMI links. As shown in Figure~\ref{fig:hierarchical-collectives}B, after this phase each XCD holds the fully reduced result for its half of the tile---aggregated across all physical GPUs.

    \item \textbf{Intra-GPU all-gather (Phase 3):} Each physical GPU performs a local all-gather across its XCDs over USR links. As shown in Figure~\ref{fig:hierarchical-collectives}C, this distributes the complete $M \times N$ result to all XCDs. Like Phase 1, all physical GPUs execute in parallel.
\end{enumerate}

\paragraph{Benefits} This decomposition provides two key advantages. First, intra-GPU and inter-GPU communication are fully decoupled: the fast USR links handle Phases 1 and 3 without being throttled by slower XGMI traffic. Second, the inter-GPU all-reduce in Phase 2 operates over physical GPUs only (e.g., 8 participants), not virtual GPUs (64 participants), reducing XGMI traffic proportionally. Together, these ensure both network tiers operate at their respective peak bandwidths.

Furthermore, because intra-GPU communication is isolated, we can tailor Phases 1 and 3 to exploit MI300X's ring topology without affecting the inter-GPU collective. This enables topology-aware optimizations---such as direction-optimized ring reduce-scatter---that would be impossible in a flat collective spanning heterogeneous links.

\subsection{Two-Dimensional GEMM Sharding}
\label{subsec:gemm}

As shown in Section~\ref{sec:motivation}, extending unidimensional sharding into partitioning creates increasingly skinny matrices with degraded arithmetic intensity. If RowParallelLinear shards along $K$ across physical GPUs, naive sharding splits further along $K$ across XCDs, producing $K/64 \times N$ shards that are severely memory-bound.

\paragraph{Orthogonal Sharding Strategy} Starscream retains the original inter-GPU sharding and selects the \textit{orthogonal} dimension for intra-GPU sharding. If physical GPUs use RowParallelLinear (sharding along $K$), XCDs use ColumnParallelLinear (sharding along $N$), and vice versa. This keeps tiles approximately square, preserving arithmetic intensity.

\paragraph{Collective Interchange} Different sharding strategies require different collectives: RowParallelLinear produces partials requiring reduce-scatter at the output, while ColumnParallelLinear requires all-gather at the input to reconstruct the full activation. Crucially, these are \textit{interchangeable}: for the same per-GPU GEMM work, both approaches produce identical output shapes---only the collective placement differs.

\begin{figure}[hb]
\vspace{-0.2cm}
\centering
\adjustbox{valign=c}{\includegraphics[width=\columnwidth]{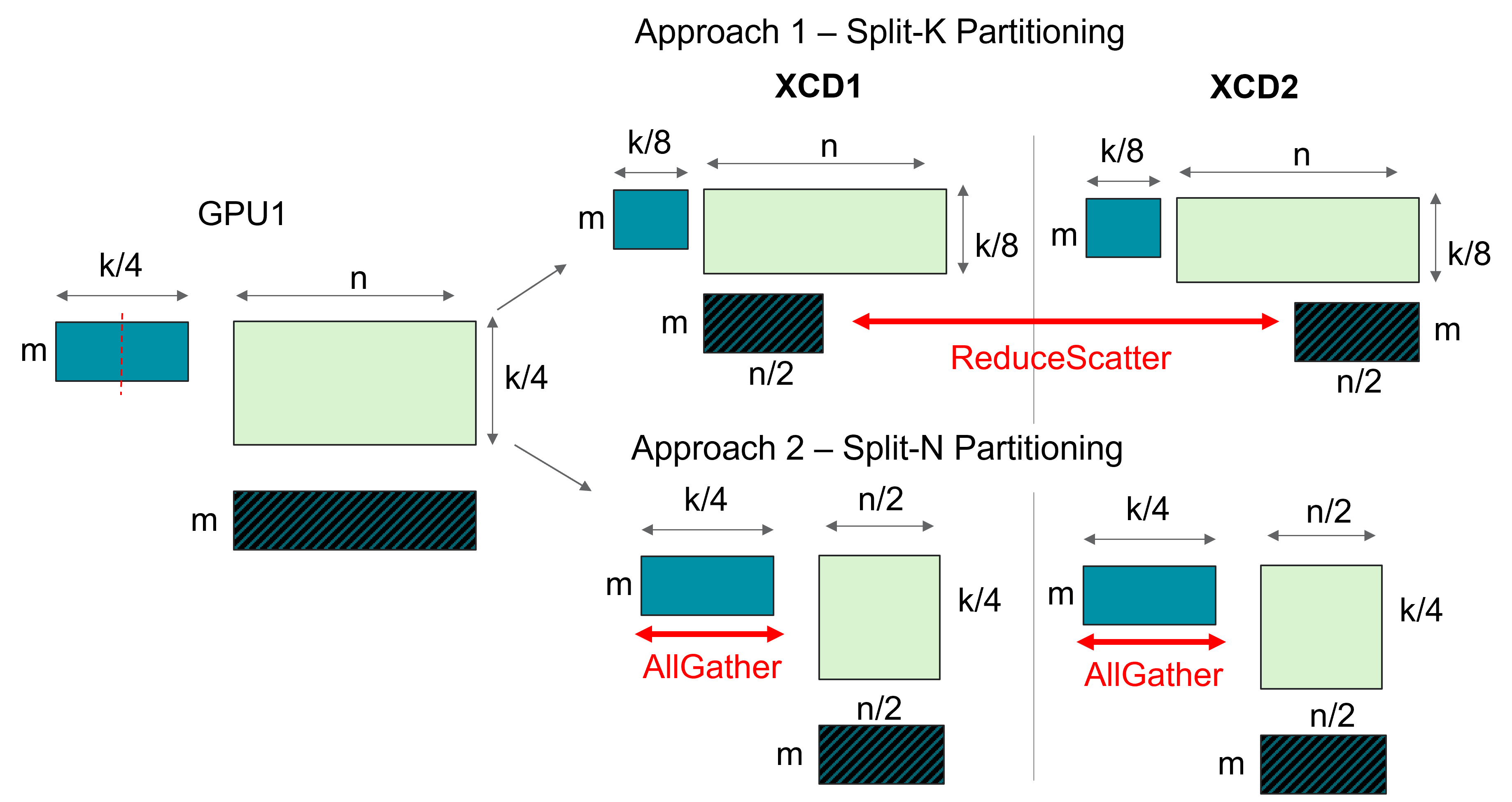}}

\hfill
\small(A)
\hspace{2cm}
\hfill
(B)
\hspace{2cm}
\hfill

\caption{\textbf{Orthogonal GEMM sharding: (A) Work allocated per GPU, (B) inter-XCD sharding: RowParallelLinear and ColumnParallelLinear.}}
\label{fig:gemm-sharding}
\end{figure}

Consider FC2 with RowParallelLinear across physical GPUs. Each physical GPU holds a $K/4 \times N$ weight shard, and each XCD receives input of shape $M \times K/8$ from the previous layer. Two options exist for intra-GPU sharding (Figure~\ref{fig:gemm-sharding}):

\begin{enumerate}
    \item \textbf{Extend RowParallelLinear:} Shard weights further along $K$, giving each XCD a $K/8 \times N$ shard. Each XCD computes $M \times N$ partials, followed by inter-XCD reduce-scatter, yielding $M \times N/2$ output per XCD.

    \item \textbf{Apply ColumnParallelLinear:} Shard weights along $N$, giving each XCD a $K/4 \times N/2$ shard. First all-gather the input to $M \times K/4$, then compute the GEMM, yielding $M \times N/2$ output per XCD.
\end{enumerate}

Both produce identical $M \times N/2$ outputs per XCD. Option 1 creates skinny $K/8 \times N$ tiles; Option 2 creates balanced $K/4 \times N/2$ tiles with 2$\times$ better arithmetic intensity.

\begin{figure*}[t]
\centering
\adjustbox{valign=c}{\includegraphics[width=0.78\linewidth]{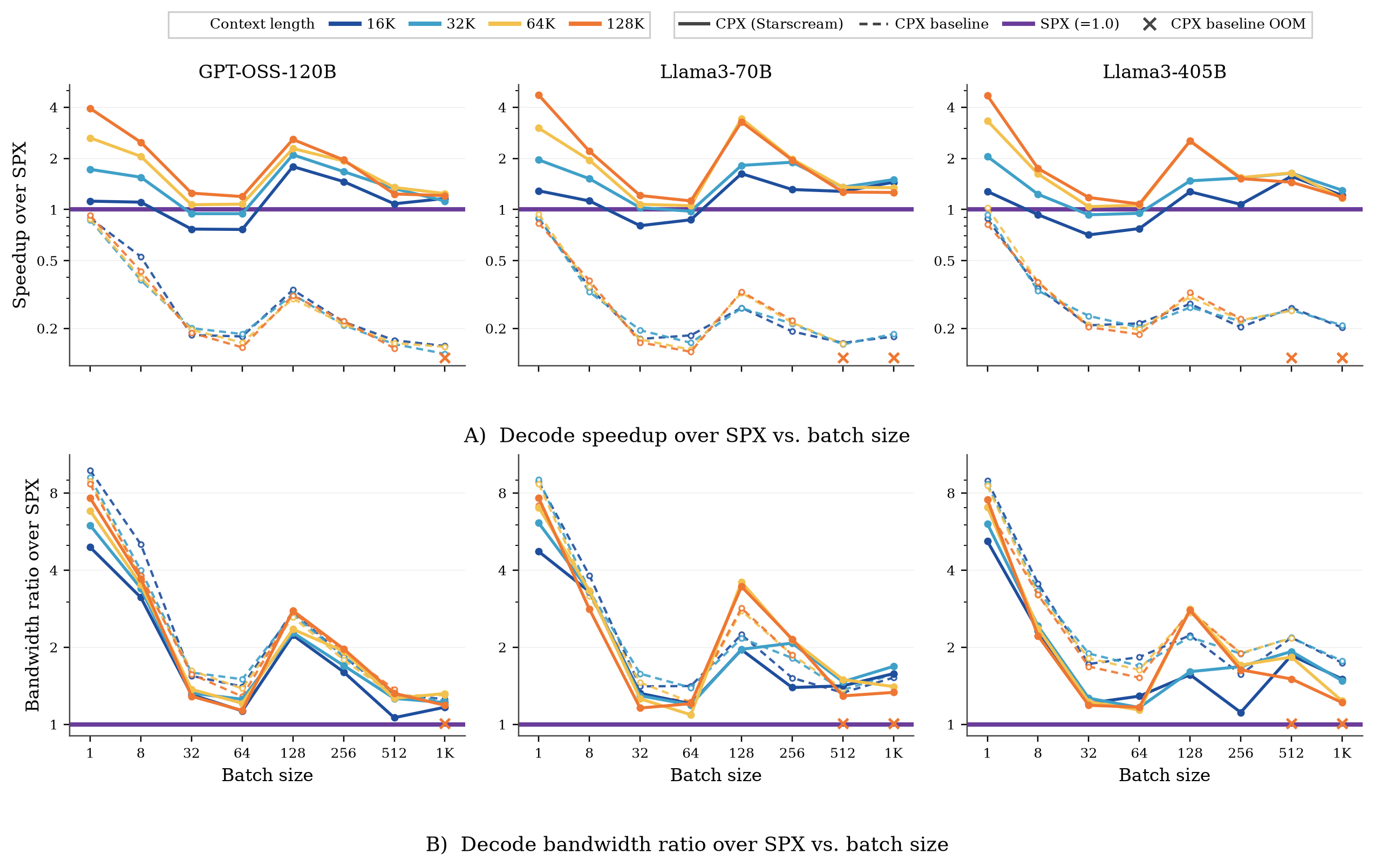}}
\hfill
\adjustbox{valign=c}{\includegraphics[width=0.2\linewidth]{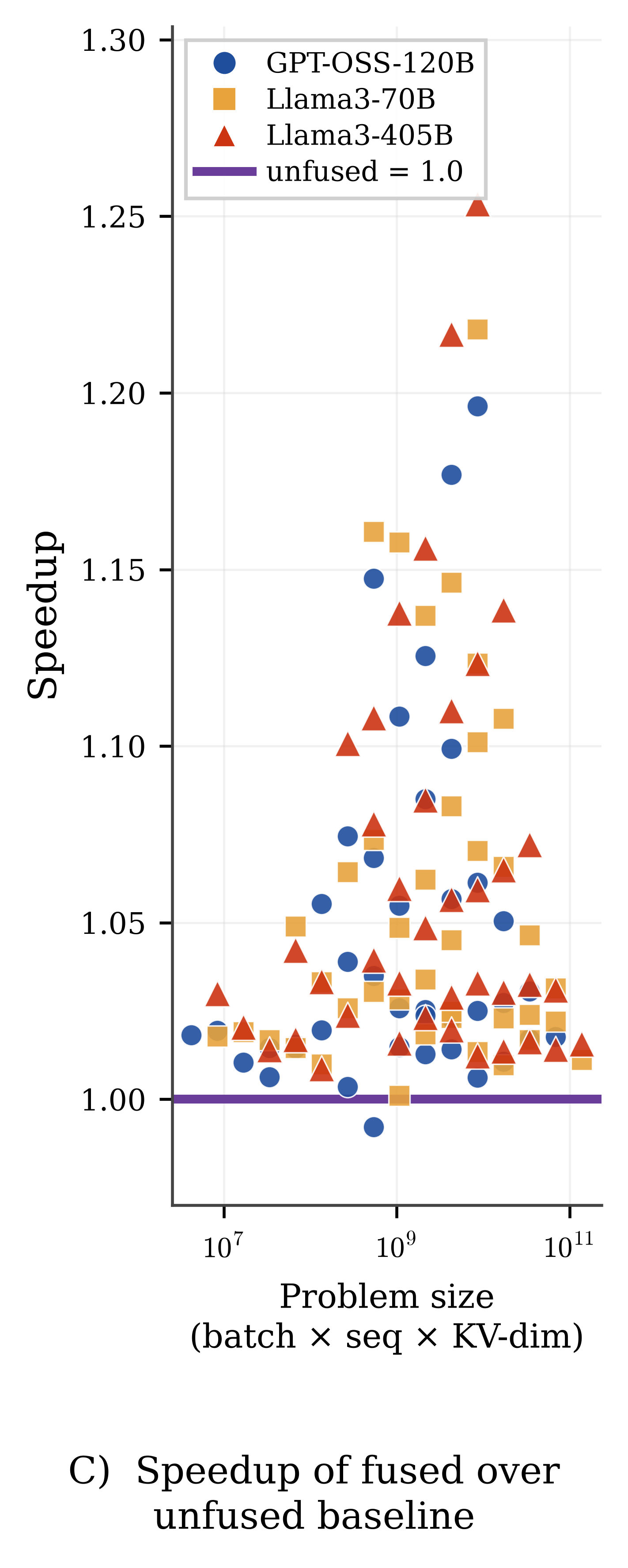}}
\vspace{-0.2cm}
\caption{ \textbf{(A) Normalized Speedup, (B) memory bandwidth utilization, (C) Additional Speedup from compute-communication overlap for GPT-OSS-120B, Llama3-70B and Llama3-405B models with varying batch size and context lengths. }}
\label{fig:attention-speedup-bandwidth}
\vspace{-0.5cm}
\end{figure*}

\paragraph{Applying to the Full LLM} In standard tensor parallelism, all-reduce boundaries occur between output projection and FC1, and between FC2 and the next layer's input projection. Output projection and FC2 use RowParallelLinear; FC1 and input projection use ColumnParallelLinear. Replacing all-reduce with hierarchical collectives (Section~\ref{subsec:collectives}), we observe: RowParallelLinear GEMMs are followed by inter-XCD reduce-scatter, and ColumnParallelLinear GEMMs are preceded by inter-XCD all-gather.

Using the collective interchange property, we can substitute any ``GEMM $\rightarrow$ reduce-scatter'' with ``all-gather $\rightarrow$ GEMM''---converting RowParallelLinear to ColumnParallelLinear at the XCD level, or vice versa. This flexibility allows \textit{all} GEMMs in an LLM layer to be 2D-sharded: orthogonal dimensions across physical GPUs and XCDs, maintaining balanced tiles and high arithmetic intensity throughout.

\subsection{Computation-Communication Overlap}
\label{subsec:fusion}

Even with hierarchical collectives and 2D sharding, discrete kernel boundaries impose three performance taxes~\cite{trifan-taxes}: (1) HBM write at compute kernel completion and read at collective kernel start, (2) bulk synchronization across all workgroups within each kernel, and (3) kernel launch overhead for the collective. These taxes compound across the many compute-collective pairs in an LLM layer.

Starscream eliminates these taxes by fusing compute and collective operations. The key insight: if a kernel's output must reside in a specific NUMA region for the subsequent kernel, we customize the compute kernel to write directly to that region as outputs are generated. Each thread block writes to the appropriate memory partition using symmetric memory~\cite{torch-symm-mem}, avoiding a separate collective kernel entirely.

We apply this fusion throughout the LLM pipeline---GEMMs, attention, RMSNorm, and any operation requiring cross-XCD communication under our sharding strategy. Whether the collective precedes (all-gather) or follows (reduce-scatter) the computation, we use software pipelining, prefetching, and careful data orchestration to overlap communication with compute, hiding whatever latency remains after hierarchical mapping.

\section{Evaluation}
\label{sec:results}

\subsection{Methodology}
\paragraph{Hardware} We evaluate Starscream on AMD Instinct MI300X and MI355X GPUs. For MI300X, we compare CPX/NPS4 mode against baseline SPX/NPS1 mode. For MI355X, we compare DPX/NPS2 mode against SPX/NPS1.

\paragraph{Implementation} Self-attention kernels build on vLLM's~\cite{vllm} Triton-based unified-attention implementation (evaluated on vLLM-v1) and the HIP-based~\cite{kwon2023efficient} paged-attention implementation (evaluated on vLLM-v0). GEMM kernels are implemented using the Stream-K~\cite{stream-k} algorithm in Triton. Computation-communication fusion leverages PyTorch's symmetric memory~\cite{torch-symm-mem}. We plan to open-source all implementations upon acceptance.

\paragraph{Workloads} For self-attention, we evaluate the decode phase for three representative model shapes---GPT-OSS-120B, Llama3-70B, and Llama3-405B---across context lengths from 16K to 128K tokens and batch sizes from 1 to 1024. For GEMM and collective kernels, we use over 200 representative shapes drawn from state-of-the-art LLMs including GPT-OSS-120B, Llama3-70B, Llama3-405B, and others. We deliberately selected shapes representing both state-of-the-art GQA-based and MoE-based architectures to ensure broad coverage.

\begin{figure*}[t]
\centering
\adjustbox{valign=c}{\includegraphics[width=0.49\linewidth]{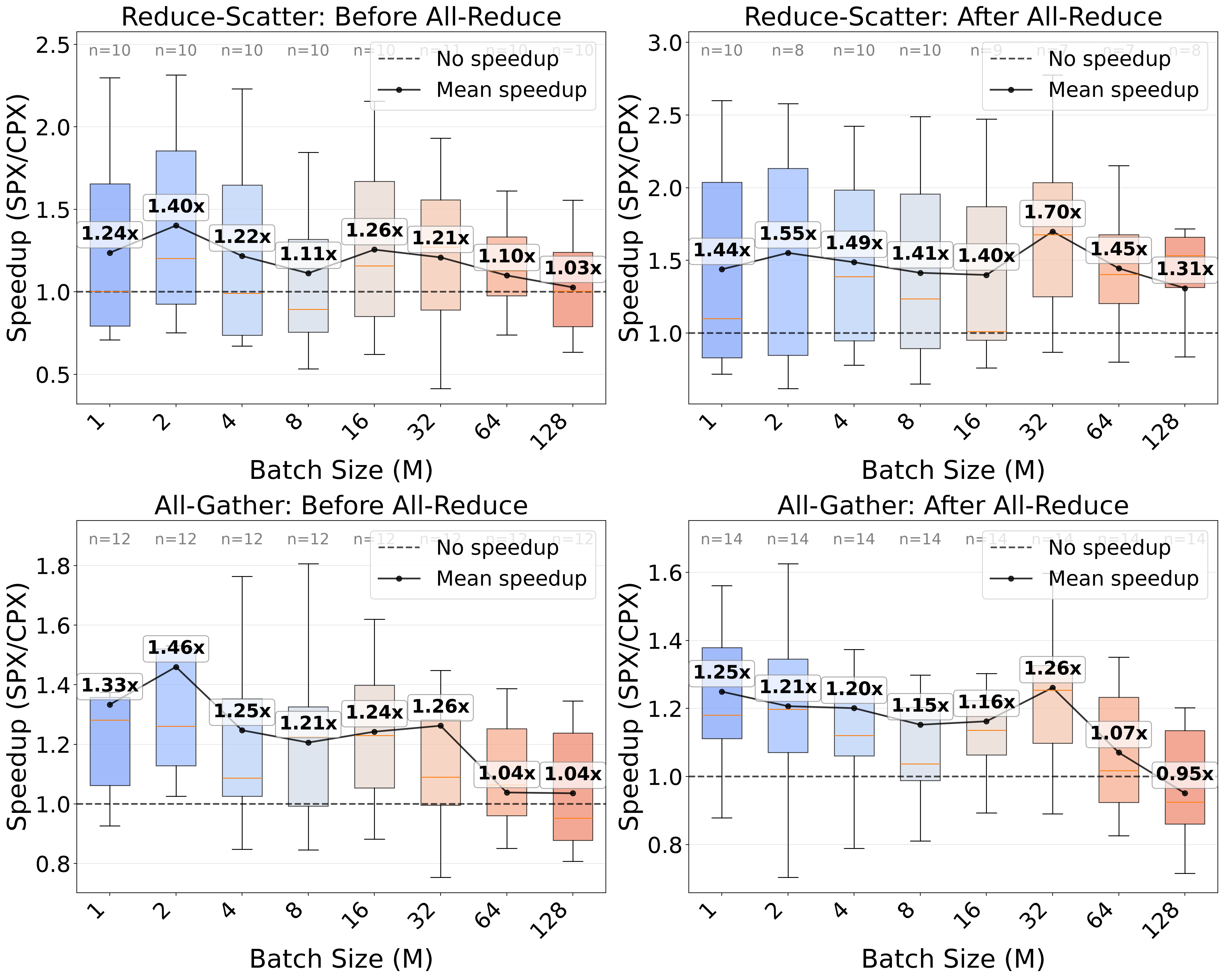}}
\hfill
\vrule
\hfill
\adjustbox{valign=c}{\includegraphics[width=0.49\linewidth]{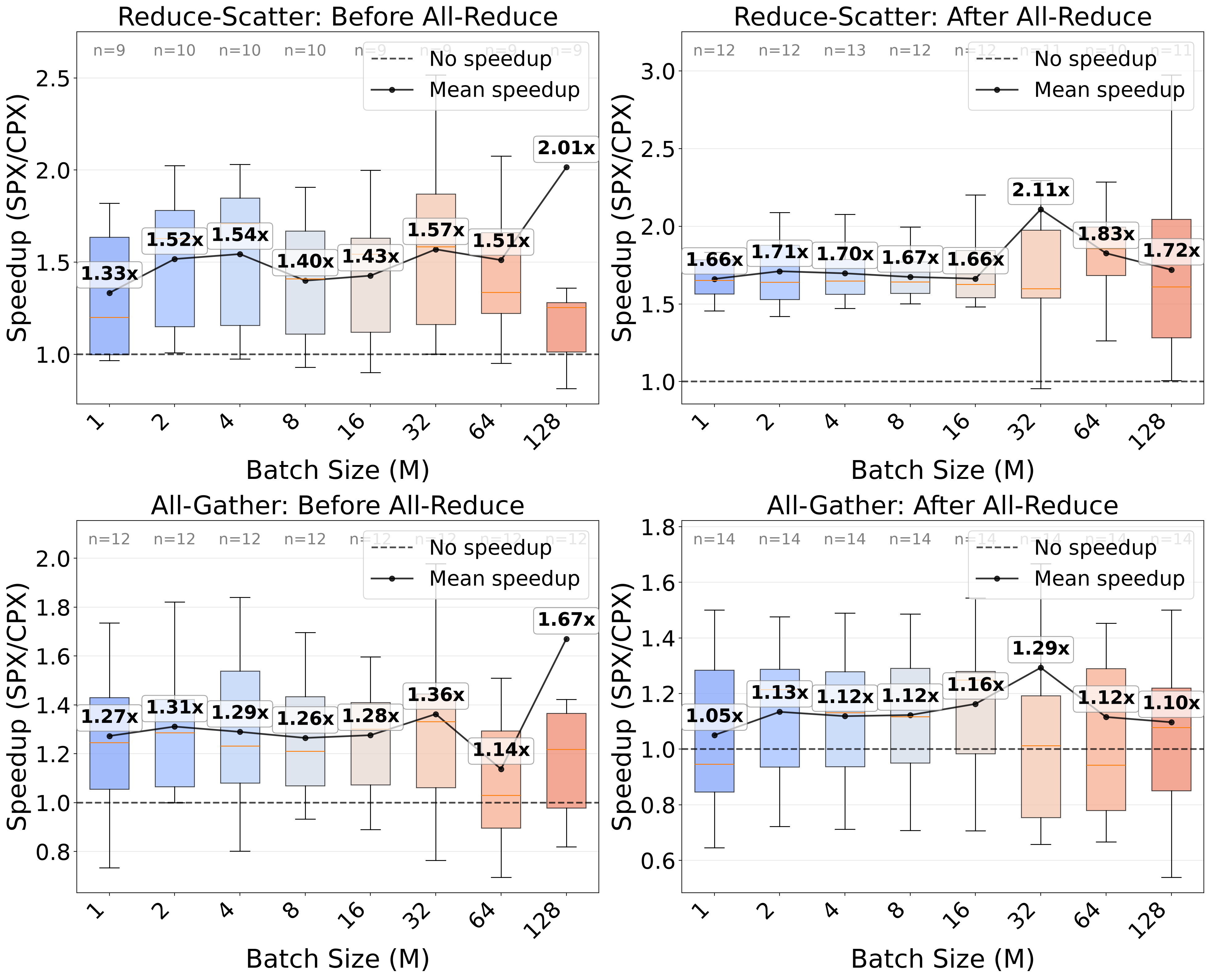}}

\hspace{4.5cm}
\small(A)
\hfill
\hspace{4.5cm}
(B)
\hfill
\vspace{-0.2cm}
\caption{\textbf{GEMMs with intra-GPU collective speedup on (A) MI300X in CPX NPS4, and (B) MI355X in DPX NPS2 against GEMMs in SPX NPS1.} }
\label{fig:gemm+collec}
\vspace{-0.3cm}
\end{figure*}

\subsection{Self-Attention Performance}

Figure~\ref{fig:attention-speedup-bandwidth}A shows the speedup of self-attention kernels during the decode phase using Starscream's sequence-length sharding (CPX Starscream) and naive sharding (CPX baseline) in CPX/NPS4 mode relative to the SPX/NPS1 baseline. Figure~\ref{fig:attention-speedup-bandwidth}B shows the corresponding HBM bandwidth utilization for the same configurations.

\paragraph{Low-Latency Scenarios} At small batch sizes, Starscream achieves its highest speedups. In SPX/NPS1 mode, small batches create memory access hotspotting---requests concentrate on a subset of HBM slices, leaving aggregate bandwidth underutilized. Starscream's sequence-length sharding distributes the KV cache evenly across NUMA domains, eliminating hotspots and saturating available bandwidth. This translates directly to improved throughput.

\paragraph{High-Throughput Scenarios} As batch sizes increase, the performance gap between Starscream and SPX/NPS1 narrows. Larger batches naturally distribute memory accesses more uniformly, reducing hotspotting even in SPX mode. Figure~\ref{fig:attention-speedup-bandwidth}B confirms this: SPX bandwidth utilization improves with batch size, approaching Starscream's utilization at the largest batches (reflected in the smaller SPX-CPX gap).

However, SPX/NPS1 at high batch sizes introduces a different bottleneck: USR link contention. With more data flowing across IODs, XCDs compete for shared interconnect bandwidth, reducing net effective bandwidth. Starscream sidesteps this by minimizing USR traffic---all-gathering the smaller tensor (queries or KV heads, depending on phase) keeps cross-die communication lightweight, sustaining performance gains even when hotspotting is no longer the dominant factor.

\paragraph{Naive Sharding Fails} The CPX/NPS4 baseline without Starscream (naive sharding) performs consistently worse than SPX/NPS1, and degrades further as batch size increases. KV cache replication is the culprit: larger batches require more KV entries, and replicating them across XCDs consumes memory bandwidth bringing redundant data. The red $\times$ markers in Figure~\ref{fig:attention-speedup-bandwidth} indicate out-of-memory failures, which occurs only with naive sharding. Naive sharding exhausts capacity at batch sizes where both SPX/NPS1 and Starscream continue operating, a direct consequence of KV cache replication.

\paragraph{Computation-Communication Overlap} As detailed in Section~\ref{subsec:fusion}, we overlap computation with communication by customizing compute kernels to write outputs directly to the target NUMA region, eliminating separate collective kernels and their associated taxes. Figure~\ref{fig:attention-speedup-bandwidth}C shows the performance benefit of this overlap over discrete kernels. This speedup compounds with Starscream's other optimizations.

\paragraph{Conservative Estimates} These results represent conservative lower bounds. We took the best-performing SPX/NPS1 kernel configuration and ported it to CPX/NPS4 with Starscream's collectives and aggregation kernels, without tuning workgroup distribution, workgroup sizes, or flash-attention segmentation for the CPX configuration. The occasional performance dips at certain batch sizes (e.g., 32, 64) reflect this lack of CPX-specific tuning. With dedicated optimization, we anticipate further speedups.

\subsection{GEMM with Intra-GPU Communication}

Figure~\ref{fig:gemm+collec} shows the performance of GEMM with intra-GPU communication on MI300X (CPX/NPS4) and MI355X (DPX/NPS2) as speedup over the baseline SPX/NPS1. Each figure contains four quadrants: the top row shows GEMM followed by reduce-scatter; the bottom row shows all-gather followed by GEMM. The first column covers GEMM shapes occurring \textit{before} the all-reduce in baseline tensor-parallel inference; the second column covers shapes occurring \textit{after} the all-reduce.

\paragraph{Connecting to Hierarchical Collectives} Recall from Section~\ref{subsec:collectives} that Starscream decomposes all-reduce into intra-GPU reduce-scatter, inter-GPU peer-XCD all-reduce, and intra-GPU all-gather. In
  SPX mode, the computation is simply GEMM $\rightarrow$ inter-GPU all-reduce $\rightarrow$ GEMM. In CPX mode, this
  becomes GEMM $\rightarrow$ hierarchical collective (reduce-scatter + all-reduce + all-gather) $\rightarrow$ GEMM.
  Since the inter-GPU all-reduce transfers the same amount of data in both cases, we isolate the additional CPX
  overhead by comparing: (1) GEMM-only in SPX against GEMM fused with intra-GPU reduce-scatter in CPX (top-left
  quadrant), and (2) GEMM-only in SPX against intra-GPU all-gather fused with GEMM in CPX (bottom-right quadrant).

\paragraph{2D Sharding via Collective Interchange} As discussed in Section~\ref{subsec:gemm}, swapping GEMM+reduce-scatter with all-gather+GEMM (and vice versa) enables 2D sharding with improved arithmetic intensity. The remaining two quadrants (top-right, bottom-left) show this alternative. Comparing across quadrants, all four variants outperform SPX/NPS1, but the 2D-sharded configurations (swapped collectives) achieve additional speedup over their 1D counterparts---purely from better tile shapes and improved arithmetic intensity.

\begin{figure}[t]
\centering
\adjustbox{valign=c}{\includegraphics[width=\columnwidth]{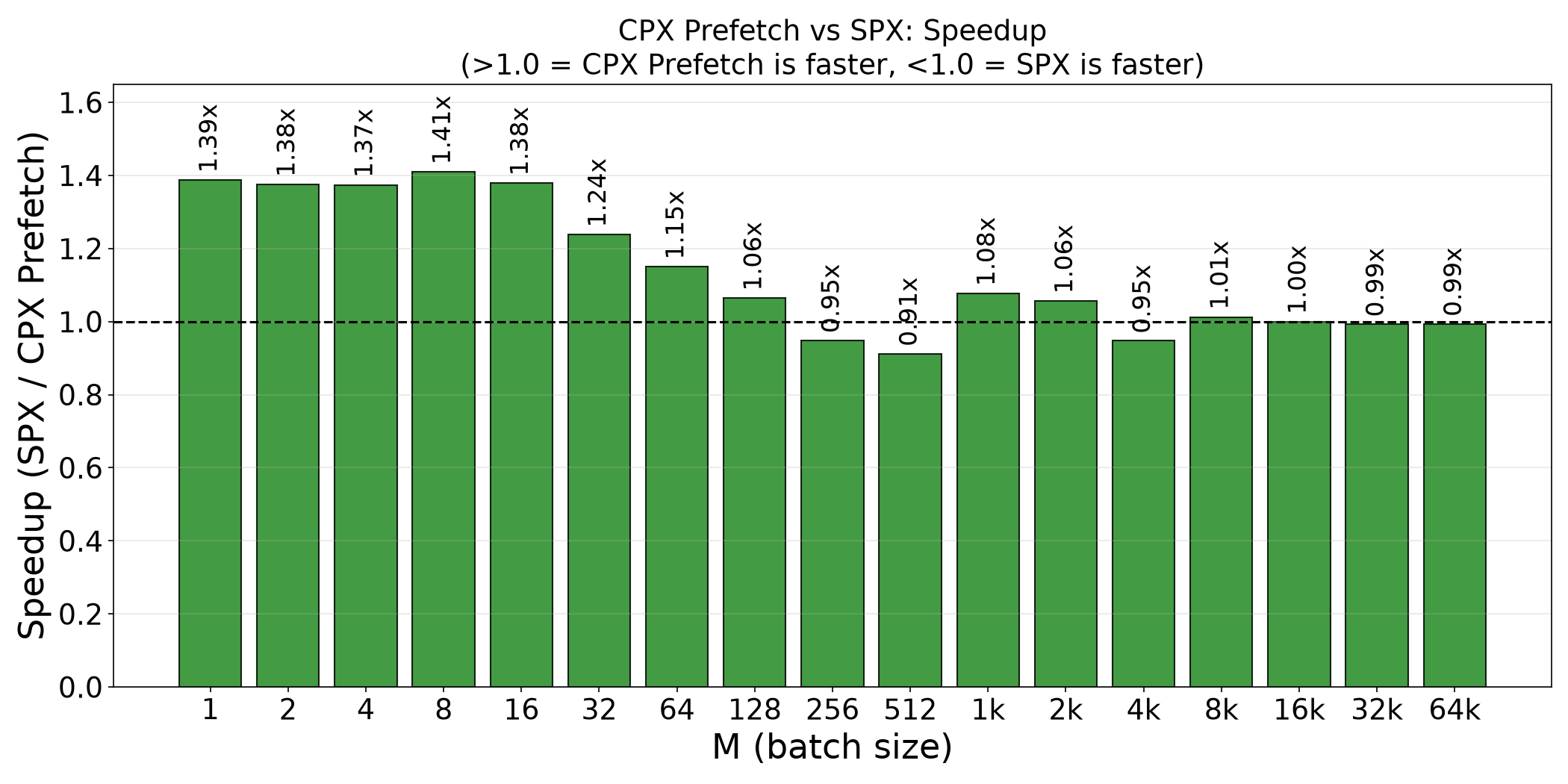}}
\caption{\textbf{Speedup of Llama3-405B in CPX NPS4 with AllGather followed by GEMM against SPX NPS1 as batch size increases.}}
\label{fig:gemmvsbatch}
\end{figure}

\paragraph{Performance Across Batch Sizes} Figure~\ref{fig:gemmvsbatch} breaks down speedup by batch size for the all-gather+GEMM configuration on Llama3-405B FC2. Performance characteristics fall into three regimes:

\begin{itemize}
    \item \textbf{Low-latency (batch 1--16):} In SPX/NPS1, all three matrices are distributed across 8 XCDs, and latency cannot be hidden. CPX/NPS4 leverages near-data compute with minimal remote access, achieving up to 1.41$\times$ speedup.                                                                                                           
    \item \textbf{Throughput-optimized (batch 32--2048):} Mid-range batches become compute-bound, allowing SPX to begin hiding memory latency. The performance gap narrows to approximately 1.08$\times$, though CPX retains an advantage from superior bandwidth utilization.

    \item \textbf{Extreme-scale (batch $>$2048):} Large batches exceed L2 capacity per XCD (FP16 on MI300X), causing sustained cache thrashing. Gains diminish at this scale; further improvement requires lower precision or larger L2 capacity.
\end{itemize}

\section{Related Work}
\label{sec:related}

\paragraph{Chiplet GPU Coherence}
  Prior work addresses coherence overhead in multi-chiplet GPUs assuming a shared
  coherence domain. HMG~\cite{hmg-hpca20} extends coherence protocols hierarchically;
  CARVE~\cite{carve-micro18} caches remote GPU data locally; CPElide~\cite{cpelide-micro24}
  tracks inter-kernel dependencies to synchronize only when necessary, further improving
  performance.

 Starscream differs fundamentally: AMD's CPX mode partitions a GPU into independent
  virtual GPUs with \textit{private} L2 caches and \textit{disjoint} memory, eliminating coherence
  overhead entirely.
  This introduces new challenges---reduced per-partition capacity and explicit
  communication---that Starscream addresses via orthogonal
  sharding. Even with zero coherence overhead, partitioning's NUMA locality benefits (faster local HBM,
  3$\times$ lower kernel launch latency) remain inaccessible to unified-memory designs like SPX NPS1.

 \paragraph{Parallelism Strategies}
 Distributing LLMs across GPUs relies on established techniques: tensor
  parallelism (TP) shards weight matrices with all-reduce reconciliation~\cite{megatron-lm};
  sequence parallelism (SP) shards activations along the sequence dimension
  for normalization and dropout~\cite{sequence-parallel, seq2, seq3}; context
  parallelism (CP) shards input sequences, exchanging KV tensors for full-context
  attention~\cite{context-parallel}; pipeline parallelism shards across
  layers~\cite{pipeline-parallel}; expert parallelism routes tokens to sharded
  MoE experts~\cite{expert-parallel}. Long-context LLMs have driven further
  refinements: Ring Attention~\cite{ring-attention} overlaps KV transfers with
  blockwise attention; Striped Attention~\cite{striped-attention} improves load
  balance via interleaved tokens; DeepSpeed-Ulysses~\cite{ulysses} uses all-to-all
  for head-parallel attention; USP~\cite{usp} and LoongTrain~\cite{loongtrain}
  unify these into hybrid schemes;
  DualPipe~\cite{deepseek-v3} overlaps forward and backward computation with
  communication to minimize pipeline bubbles. 
  Helix~\cite{helix} goes further, observing that a single uniform sharding is
  suboptimal: it disaggregates attention and FFN \textit{in time}, reconfiguring the
  same GPUs between KV-parallel attention and TP FFN phases, bridged by
  a single all-to-all across a flat NVLink domain.

  These approaches share a common assumption: a homogeneous inter-GPU fabric.
  Starscream exploits a different axis of heterogeneity---the \textit{intra-GPU}
  bandwidth hierarchy exposed by chiplet partitioning. Rather than replacing
  inter-GPU parallelism, Starscream composes \textit{in space}: it retains the
  inter-GPU layout unchanged and adds an orthogonal intra-die dimension across
  XCDs, exploiting the 10$\times$ bandwidth gap between on-package USR and inter-die
  XGMI. Starscream further reinterprets parallelism through \emph{resource reduction}:
  partitioning trades aggregate resources per virtual GPU for locality, and the
  optimal composition under this regime differs from the flat-topology optimum.
  This drives our design choices: where Helix and Ring Attention use uniform
  collectives, Starscream employs hierarchical collectives tuned to the NUMA
  topology and selects query- or KV-side movement per inference phase. The approaches
  are complementary: any inter-GPU strategy---Ring, Ulysses, or Helix---can serve
  as the outer layer that Starscream augments within each physical GPU.

\paragraph{Partitioned and spatial accelerators}
Beyond GPUs, spatial and wafer-scale accelerators expose fine-grained control
over compute and data placement. Cerebras maps models onto a wafer-scale array
of cores with software-managed on-chip data movement~\cite{cerebras}, and a
broad body of accelerator work explores tiling and dataflow mappings across
spatially partitioned compute~\cite{candles, wax, centaur, isaac, simba, gemini, waferscale-2048, nn-baton, siam, musavi-comm-char, claire, mozart, watos}. These systems demonstrate the
value of locality-aware mapping, but they do so for bespoke architectures with
programming models distinct from the GPU software stack. Starscream shows that
comparable locality benefits are attainable \emph{within} the commodity GPU
programming model, by exploiting hardware partitioning modes rather than
requiring a new class of accelerator.


\section{Conclusion}
\label{sec:conclusion}

Non-uniform memory access is an inherent reality in modern GPUs—whether from larger monolithic dies or chiplet-based designs.
  AMD's compute and memory partitioning modes expose this topology, providing knobs to exploit the underlying NUMA
  hierarchy. With perfect isolation, partitioning delivers its full potential. But when a model exceeds a single partition's capacity,
  efficient mapping is essential to maximize reuse and minimize communication; without it, redundant parameters, KV caches, skinny
  matrices, and flat collectives erode the architectural gains. 
In this work, we presented Starscream, which reframes partitioning not as ``more GPUs'' but as a new level of the compute and memory hierarchy, by composing existing parallelism strategies hierarchically and applying each at the level of the topology where it is effective. 
Concretely, Starscream eliminates redundant memory usage, sustains arithmetic intensity through 2D GEMM sharding, structures collectives around the intra-die/inter-die bandwidth gap, and fuses computation and communication to further improve performance. 
On AMD Instinct MI300X and MI355X GPUs, these techniques deliver up to 4$\times$ faster attention and 1.43$\times$ faster GEMM execution, transforming GPU partitioning into a net advantage—using the same compute resources, with no additional hardware. 
As chiplet counts continue to grow and NUMA effects deepen, Starscream's hierarchical sharding principles offer a blueprint for extracting performance from the topology of future partitioned accelerators.

\bibliographystyle{IEEEtranS}
\bibliography{refs}

\end{document}